\documentclass[lettersize,journal]{IEEEtran}
\usepackage{amsmath,amsfonts}
\usepackage{array}
\usepackage{textcomp}
\usepackage{stfloats}
\usepackage{url}
\usepackage{verbatim}
\usepackage{graphicx}
\usepackage{cite}

\usepackage{multirow}
\usepackage{adjustbox}
\usepackage[table,xcdraw]{xcolor}
\usepackage[ruled,linesnumbered]{algorithm2e}

\usepackage[normalem]{ulem}

\begin{document}

\title{Lossless Compression of Lookup Tables for Hardware Applications}

\author{\IEEEauthorblockN{Alireza Khataei, Kia Bazargan}\\
\IEEEauthorblockA{Department of Electrical and Computer Engineering\\
University of Minnesota\\
Minneapolis, MN, USA\\
\{khata014, kia\}@umn.edu}}

\maketitle

\begin{center}

\footnotesize

This work has been submitted to the IEEE for possible publication. Copyright may be transferred without notice, after which this version may no longer be accessible.

\end{center}

\begin{abstract}
\textcolor{black}{
Large lookup tables are widely used in hardware to store constant-valued arrays for applications ranging from elementary mathematical operations, such as constant-coefficient multiplication and nonlinear function evaluation, to emerging machine learning models, including table-based neural networks (NNs) and Kolmogorov–Arnold networks (KANs). However, storing extensive tables of constant values can lead to excessive hardware costs in resource-constrained edge devices such as FPGAs. In this paper, we propose CompressedLUT, a lossless compression scheme and its decoder hardware architecture for the efficient storage and retrieval of arbitrary data in hardware. Our method combines decomposition, self-similarities, higher-bit compression, and multilevel compression techniques to maximize table size savings without accuracy loss. Its hardware decoder primarily uses addition, arithmetic right shift, and several small lookup tables, ensuring low area and high throughput. We evaluated CompressedLUT on FPGAs by implementing multiple nonlinear functions, constant-coefficient multipliers (CCMs), and KANs at 12-bit resolution. CompressedLUT is available as an open-source tool.}
\end{abstract}


\section{Introduction}

Lookup tables are widely used in both hardware and software systems to store blocks of read-only, predefined data. Such tables are used in field-programmable gate arrays (FPGAs), graphics processing units (GPUs), and digital signal processors (DSPs). Their applications range from elementary mathematical operations, such as constant coefficient multipliers (CCMs) and nonlinear functions, to novel machine learning models, such as table-based neural networks (NNs) and Kolmogorov–Arnold networks (KANs). Compressing lookup tables can potentially reduce their implementation costs in terms of memory resource utilization, throughput, power consumption, \textit{etc.} This issue has been of considerable interest as an active research area~\cite{9170828, 6998028, 9628172, 8920330, 8350933, 10.1145/3626202.3637575, 10.1145/3706628.3708823}.

\textcolor{black}{Nonlinear functions have various applications such as activation functions in machine learning.} Using lookup tables for function evaluation is an efficient method due to its simplicity of implementation, low computational latency, and high throughput, especially for evaluating compound complex functions, such as $1/[1+e^{-x}]$, which can be evaluated by a table of precomputed values in hardware instead of performing costly intermediate operations step-by-step. At low resolutions, lookup tables can be directly used for implementing a function by tabulating the values of all possible inputs. Given a function at the input resolution $w_{in}$ and output resolution $w_{out}$, the size of the corresponding lookup table would be $w_{out}~\times~2^{w_{in}}$ bits, which grows exponentially as $w_{in}$ increases. For this reason, this approach is usually used for evaluating a function at up to 12-bit resolutions~\cite{9106347}. At higher resolutions, however, simple tabulation of a function is not feasible due to the massive sizes of resulting tables. In such cases, approximate methods are applied, which sacrifice accuracy for hardware cost savings. Examples of such methods include bipartite table (BT)~\cite{795125}, multipartite table (MT)~\cite{1388196, 7480839}, and piecewise polynomial approximation (PPA)~\cite{1540405} methods. BT and MT decompose the table of a function into smaller tables, called the table of initial values (TIV) and table of offsets (TO), which result in the reduction of hardware costs. PPA methods, however, break a function into sub-functions and approximate them with polynomials whose coefficients are stored in smaller tables. Although all of these methods can simplify the implementation of a function at the expense of accuracy, they still rely on lookup tables to store essential values such as TIV, TO, or tables of coefficients. Lookup tables are also used in other state-of-the-art methods, libraries, and architectures for high-resolution function evaluation. For instance, hls4ml~\cite{fastml_hls4ml, Duarte:2018ite} is a Python package for machine learning inference on FPGAs, and it uses lookup tables to perform nonlinear parts of activation functions in NNs. Additionally, many floating-point operations require lookup tables as parts of their architectures~\cite{9806139, 7563286, 10.1007/978-3-030-79025-7_22}. As a result, lookup tables are used either directly or as parts of other table-based methods for function evaluation. In either case, table compression methods can be used to shrink such tables to reduce their implementation hardware costs.

\textcolor{black}{Multiplication is a fundamental operation in many applications. When one operand is constant, the operation can be implemented as a CCM using specialized hardware instead of a general-purpose multiplier~\cite{7752883, 5753874, Walters2017ReducedAreaCA, 9114822, 10.1145/3494570, 10323844}. Because a CCM computes the linear function \(y=cx\) for a fixed coefficient \(c\), it can also be implemented as a lookup table that stores the output for each possible input value. In this application, lookup tables contain many redundancies that compression methods can exploit to reduce hardware costs.}

\textcolor{black}{NNs require repeated multiply-accumulate operations, nonlinear activations, and access to weights and biases, all of which contribute to hardware costs. Recently, table-based methods~\cite{10.1145/3748173.3779200, 11552695, 11008982, 8735521, 9221584, 10416099, 10705559, 10705569} have been proposed in which a neuron or a sub-network is represented as a lookup table that directly maps the inputs to an output. These methods illustrate another use of lookup tables in hardware. Recently, ReducedLUT~\cite{10.1145/3706628.3708823} extended CompressedLUT~\cite{10.1145/3626202.3637575} by injecting don’t cares into the compression
process and demonstrated how it could improve the hardware costs of table-based NNs with minimal model accuracy degradation.}

 \textcolor{black}{KANs~\cite{ICLR2025_afaed896} are inspired by the Kolmogorov–Arnold representation theorem and considered as alternatives to multilayer perceptrons (MLPs). Whereas MLPs use fixed nonlinear activation functions at nodes, KANs place learnable nonlinear functions on edges. KANEL\'E~\cite{10.1145/3748173.3779202} and LUT-KAN~\cite{KUZNETSOV2026103923} provide table-based methods for KAN implementations in which edge functions are replaced with lookup tables, enabling low-latency, high-throughput inference. However, KANs typically use many nonlinear functions, and storing them in lookup tables can require substantial hardware resources, especially at high input resolutions. As a result, table compression methods can be beneficial in reducing hardware costs in this application.}

 In this paper, \textcolor{black}{which is the extended version of~\cite{10.1145/3626202.3637575},} we propose CompressedLUT as a method for lossless compression of lookup tables, which uses the idea of decomposition~\cite{6998028, 9628172}, self-similarities~\cite{10.1145/3543622.3573181, 10171591}, multilevel compression, and higher-bit compression to maximize table size savings. CompressedLUT is available as an open-source tool\footnote{{CompressedLUT is available at \textcolor{blue}{\textbf{\url{https://github.com/kiabuzz/CompressedLUT}}}} (DOI: 10.5281/zenodo.10431619).}.

\textcolor{black}{For evaluating our method, we use CompressedLUT in the following applications to implement their lookup tables efficiently at lower hardware costs while maintaining their baseline accuracy. The results were compared to implementations of lookup tables through the PlainTable (uncompressed plain table) and previous TwoTable (two-table decomposition)~\cite{6998028} approaches. All the designs are described in RTL, synthesized, and placed and routed on FPGAs.
 \begin{itemize}
 \item Nonlinear Functions: We target several nonlinear functions at 12-bit resolution. We develop a software script to evaluate a given function over a given domain and store its quantized results in a large lookup table for hardware implementations. In terms of compression, our method compresses the tables on average by 80\%, whereas TwoTable compresses them on average by 52\%. In terms of throughput per LUT (TPL) hardware cost, our method on average has 3.30 times higher throughput than PlainTable, whereas TwoTable on average has 2.15 times higher throughput.
 \item CCMs (Constant Coefficient Multipliers): We implement several CCMs at 12-bit resolution with different constant values. As CCMs are basically linear functions, we use the same approach as used in the nonlinear functions. In other words, we develop a software script to generate the lookup table values corresponding to the quantized result of the linear functions representing the CCMs. In terms of compression, our method compresses the tables on average by 94\%, whereas TwoTable compresses them on average by 65\%. In terms of TPL, our method has 10.55 times higher throughput than PlainTable on average, whereas Vivado (direct Verilog multiplication with \texttt{assign y = c * x}) and TwoTable have 1.61 and 2.74 times higher throughput on average, respectively.
 \item KANs (Kolmogorov-Arnold Networks): We use KANEL\'E (a table-based KAN framework)~\cite{10.1145/3748173.3779202} to train two KAN models on the MNIST and JSC-OpenML datasets using 12-bit learnable activation functions. That framework then generates lookup tables corresponding to the trained activation functions, which are finally compressed and implemented by our method. Considering the hardware costs of the entire networks, CompressedLUT improves TPL on average by 1.87 times compared to PlainTable.
  \end{itemize}}

The rest of the paper is organized as follows. Section~\ref{sec:compressedlut} discusses the details of each technique used for compression. \textcolor{black}{Section~\ref{sec:toolflow} shows the CompressedLUT tool flow.} In Section~\ref{sec:impleresults}, the implementation results are presented and discussed. Finally, the paper is concluded in Section~\ref{sec:conclusions}.

\section{Methodology}\label{sec:compressedlut}

We describe our compression methodology by first presenting the idea of breaking a table into two smaller tables (Sec.~\ref{sec:methodTableDecomp}), similar to what TwoTable~\cite{6998028} and LDTC~\cite{9628172} use. Then we use the idea of finding self-similarities in the smaller table (Sec.~\ref{sec:selfSimilarityIdea}), extending the idea in \cite{10171591}. 

The above methods would be suitable for tables that store functions that are smooth and have small local variations. However, for tables that store values with higher dynamic range and large local variations, such as the ones used in many function approximation methods, we present two other techniques detailed in sections~\ref{sec:higherBitCompIdea} and \ref{sec:multilevelCompIdea}. The overall architecture of our method is discussed in Sec.~\ref{sec:overallArch}.

\subsection{Lookup Table Decomposition}\label{sec:methodTableDecomp}

\begin{figure} [] 
	\centering
		\includegraphics[scale=.4] {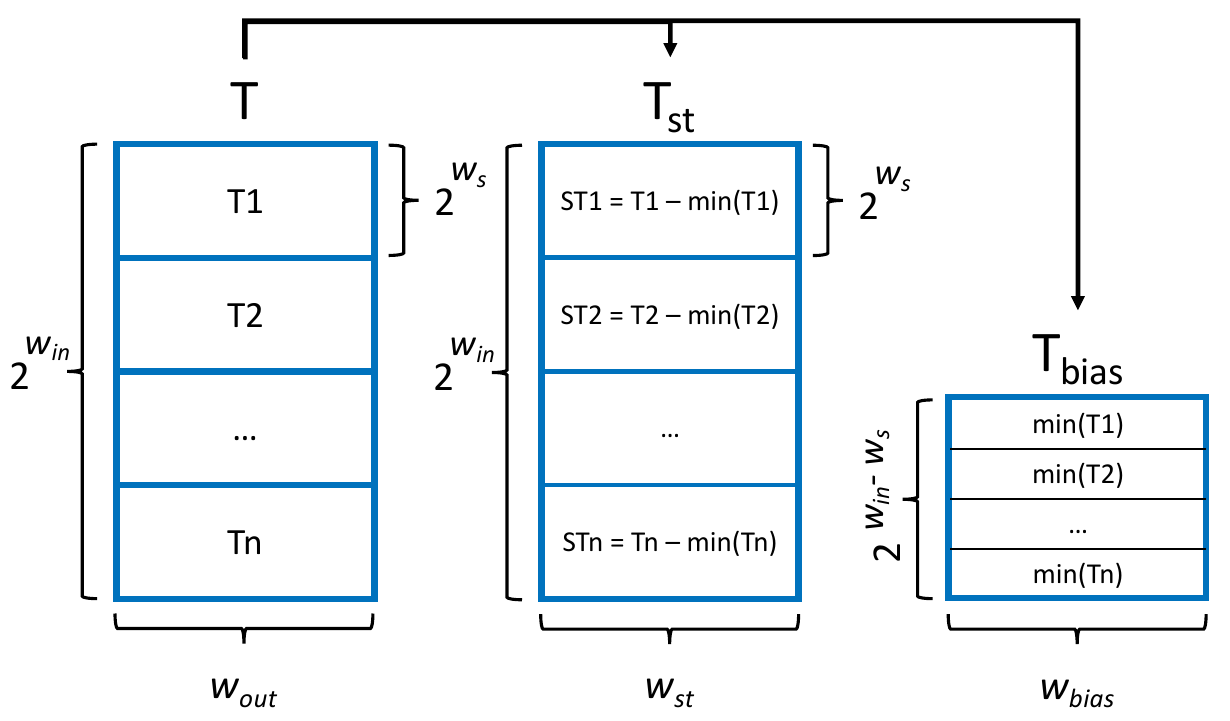}
	\caption{Decomposition of T into T\textsubscript{bias} and T\textsubscript{st}.}
 \label{fig:decomposition}
\end{figure}

Similar to TwoTable~\cite{6998028} and LDTC~\cite{9628172}, we decompose a table $T$ into two new tables $T_{bias}$ and $T_{st}$. Fig.~\ref{fig:decomposition} shows the decomposition of $T$ into $T_{bias}$ and $T_{st}$. Assuming $T$ has $2^{w_{in}}$ elements of $w_{out}$ bits, it is split into $n = 2^{w_{in}-w_s}$ sub-tables, where $0 < w_s < w_{in}$. Next, the minimum value of each sub-table is stored as an element in $T_{bias}$. Additionally, the minimum value of each sub-table is subtracted from all the values in the corresponding sub-table and the resulting values are stored in $T_{st}$.

As seen, $T_{bias}$ has $2^{w_{in}-w_s}$ elements of $w_{bias}$ bits, where $w_{bias}$ is usually the same as $w_{out}$. Whereas $T_{st}$ has $2^{w_{in}}$ elements of $w_{st}$ bits, where $w_{st}$ is less than $w_{out}$. This is because $T_{st}$ holds local variations which usually require a smaller bit width. In summary, the table $T_{bias}$ has the same output bit width as the original table $T$, but it has fewer elements. In contrast, the table $T_{st}$ has the same number of elements as the original table $T$, but it has less output bit width. The tables have the following number of bits.
\begin{equation*}
  \begin{aligned}
&Size(T) = 2^{w_{in}}\times w_{out}\\
&Size(T_{bias}) = 2^{w_{in}-w_s}\times w_{bias}\\
&Size(T_{st}) = 2^{w_{in}}\times w_{st}\\
  \end{aligned}
\end{equation*}
The final size ratio obtained by table decomposition is as follows.
\begin{equation*}
  \begin{aligned}
SizeRatio & = [Size(T_{bias})+Size(T_{st})]/{Size(T)}\\
& = [2^{w_{in}-w_s}\times w_{bias} + 2^{w_{in}}\times w_{st}]/(2^{w_{in}}\times w_{out})\\
& = 2^{-w_s} + w_{st}/w_{out}
  \end{aligned}
\end{equation*}

As seen, the final size ratio after decomposition depends on two terms: $2^{-w_s}$ and $w_{st}/w_{out}$. The parameter $w_s$ can be set to any value between 0 and $w_{in}$. Increasing $w_s$ decreases the first term $2^{-w_s}$, yet it increases the second term $w_{st}/w_{out}$. This is because increasing $w_s$ results in sub-tables with more elements, which might have larger local variations, that require greater bit width $w_{st}$.

\begin{figure} [] 
	\centering
		\includegraphics[scale=.45] {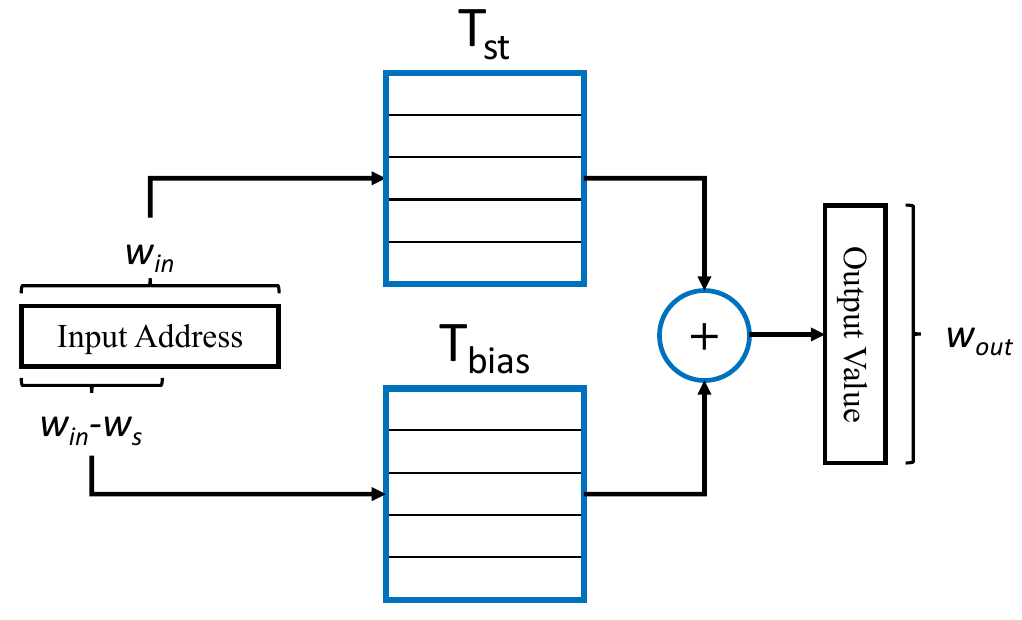}
	\caption{Retrieving T through T\textsubscript{bias} and T\textsubscript{st}.}
 \label{fig:decoder}
\end{figure}

After decomposition, the original table $T$ is replaced by $T_{bias}$ and $T_{st}$. The input address of $T_{st}$ is the same as the input address of $T$, but the input address of $T_{bias}$ is fed by the $({w_{in}-w_s})$ higher bits of the input address of $T$. Finally, an adder is used to retrieve the values of the original table $T$ by adding the output values of $T_{st}$ and $T_{bias}$, as seen in Fig.~\ref{fig:decoder}.

\subsection{Self-Similarities in Lookup Tables}\label{sec:selfSimilarityIdea}
Using the core idea of what the authors of \cite{10171591} call the ``SimBU'' method, we can compress the table of $T_{st}$ further. SimBU was proposed in the context of ``unary'' methods to reduce the complexity of HBU~\cite{10.1145/3287624.3287706}. However, the self-similarity algorithm proposed by this method can be deployed as a lossless compression approach in the context of binary lookup tables.

As discussed in Section~\ref{sec:methodTableDecomp}, $T_{st}$ holds the values of $n$ sub-tables $ST_i$, where $i\in\{1, 2, \cdots, n\}$. However, the values of many of these sub-tables are similar. Similar sub-tables refer to the sub-tables whose values are either identical or can get identical through the arithmetic right shift operation.

\begin{figure} [] 
	\centering
		\includegraphics[scale=.46] {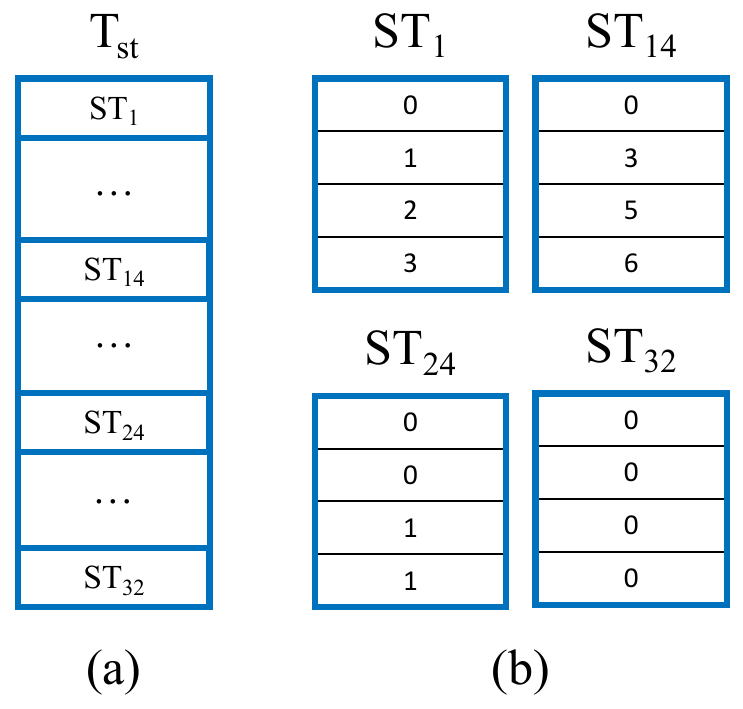}
	\caption{Examples of self-similarities in sub-tables.}
 \label{fig:similarity}
\end{figure}

Fig.~\ref{fig:similarity}a shows an example of $T_{st}$ which contains 32 sub-tables of 4 elements. Therefore, $T_{st}$ has 128 elements in total. The values of four sub-tables $ST_{1}$, $ST_{14}$, $ST_{24}$, and $ST_{32}$ are shown separately in Fig.~\ref{fig:similarity}b.  As seen, if the values of $ST_{14}$ are shifted to the right by 1 bit, we can obtain $ST_1$. Additionally, if the values of $ST_{14}$ are shifted to the right by 2 or 3 bits, we can obtain $ST_{24}$ or $ST_{32}$, respectively. In other words, we can say that $ST_{14}$ can generate $ST_1$, $ST_{24}$, and $ST_{32}$ using the right shift operation. As a result, instead of storing 4 different sub-tables, we can store only $ST_{14}$ as a unique sub-table, through which we can retrieve the other ones.

In the example of Fig.~\ref{fig:similarity}b, in addition to $ST_{14}$, $ST_1$ can also generate $ST_{24}$ and $ST_{32}$, but this time by getting shifted to the right by 1 and 2 bits, respectively. Furthermore, $ST_{24}$ can generate $ST_{32}$ by getting shifted to the right by 1 bit. However, among these 4 sub-functions, considering $ST_{14}$ as the unique sub-function is the best choice since it can generate 3 other sub-functions. Therefore, the final goal of this phase is to find the minimum set of unique sub-tables in $T_{st}$ that can generate the rest.

 Using the self-similarity matrix used in SimBU~\cite{10171591}, similarities among all sub-tables in $T_{st}$ are identified. As discussed in Section~\ref{sec:methodTableDecomp}, $T_{st}$ consists of $n = 2^{w_{in}-w_s}$ sub-tables, and each sub-table consists of $2^{w_s}$ elements. To measure similarities, an $n\times n$ Boolean matrix is needed, which is called a similarity matrix. Each entry of this matrix specifies whether the two sub-tables are similar or not. That is, an entry $s_{ij}$ is 1 if the sub-table $ST_i$ can generate $ST_j$. Obviously, this matrix is not symmetric since if $ST_i$ can generate $ST_j$ through right shifting, the opposite is not necessarily true. The following is the definition of the similarity matrix.
 \begin{equation}
 \resizebox{0.9\columnwidth}{!}{$
  \begin{aligned}
  \label{eq:sm}
SimilarityMatrix = \begin{bmatrix}
 sm_{1,1}&sm_{1,2}  &\cdots  & sm_{1, n}\\ 
 sm_{2,1}&  sm_{2,2} &\cdots  & sm_{2,n}\\ 
 \vdots &\vdots   &\ddots   &\vdots  \\ 
 sm_{n, 1}&sm_{n,2}  & \cdots & sm_{n,n}
\end{bmatrix}_{n\times n}\\
 sm_{i, j} = 1 \Leftrightarrow  \exists t \in \mathbb{N}: \forall m\in\mathbb{N} (m < 2^{w_s}), rsh_t\{ST_{i}[m]\} = ST_j[m]
  \end{aligned}
  $}
\end{equation}
where $rsh_t$ denotes an arithmetic right shift by $t$ bits.

After identifying similar sub-tables, the unique set of them should be determined that can generate the other sub-tables to retrieve the original $T_{st}$. Unique sub-tables are named $UST$, and they are all stored in a new single table, called $T_{ust}$. Furthermore, two new tables of $n$ elements, called $T_{idx}$ and $T_{rsh}$, are needed to retrieve the original table $T_{st}$ through $T_{ust}$. The value of the $i$th element in $T_{idx}$ shows the index of the unique sub-function that can generate $ST_i$, and the value of the $i$th element in $T_{rsh}$ shows the number of right bit shifts that need to be performed on the values of the corresponding unique sub-table to retrieve $ST_i$. For instance, if $T_{idx}[5] = 3$ and $T_{rsh}[5] = 2$, we can conclude that $ST_5$ can be retrieved by $UST_3$ after right shifting the values of $UST_3$ by 2 bits.

To find unique sub-tables, a vector must be obtained based on the similarity matrix. This vector is called a similarity vector, and the $j$th entry in it specifies how many sub-tables can be generated using the $j$th sub-table. The vector can be created by adding the values in each column in the similarity matrix as follows.
 \begin{equation}
  \begin{aligned}
SimilarityVector &= [sv_{1}, sv_2, \cdots, sv_n]\\
 sv_j &= \sum_i sm_{ij}
  \end{aligned}
\end{equation}

The index of the element in the similarity vector with the maximum value determines the first unique sub-table. In other words, if $sv_i$ is the element with the maximum value, $ST_i$ will be considered as the first unique sub-table $UST_1$, and its values are stored in $T_{ust}$. We also need to traverse through the $i$th column of the similarity matrix to see which sub-tables can be generated through $ST_i$. If $ST_i$ can generate $ST_j$ through right shifting by $t$ bits, then the $j$th element of $T_{idx}$ and $T_{rsh}$ must be set to 1 and $t$, respectively. After finding the first unique sub-table, we need to update the similarity matrix and similarity vector. Therefore, the $i$th row and column of the similarity matrix must be set to 0. Additionally, if $ST_i$ can generate $ST_j$, the $j$th row and column of the similarity matrix must be set to 0 as well. The elements of the similarity vector need to be recalculated based on the updated similarity matrix.

The process above needs to be repeated again and again until all the entries of the similarity matrix are 0's. In each iteration, it identifies a new unique sub-table. In the end, if the process takes $k$ iterations to finish, we will end up with $k$ unique sub-tables $UST_i$, where $i\in \{1, 2, \cdots, k\}$ and $k\leq n$. These unique sub-tables are all stored in $T_{ust}$. 

As a result, $T_{st}$ is replaced by $T_{ust}$, $T_{idx}$, and $T_{rsh}$. In contrast to $T_{st}$, which contains $n$ sub-tables, $T_{ust}$ contains $k$ unique sub-tables, where $k$ is often significantly less than $n$. It means that many sub-tables can be generated using a few unique sub-tables. Therefore, we can achieve significant table-size reductions. However, when calculating the overall memory space reduction, the size of $T_{idx}$ and $T_{rsh}$ must be taken into account. In summary, the size of each table and the size ratio are as follows.
\begin{equation*}
  \begin{aligned}
&Size(T_{st}) = n\times 2^{w_s}\times w_{st}\\
&Size(T_{ust}) = k\times 2^{w_s}\times w_{st}\\
&Size(T_{idx}) = n\times w_{idx}\\
&Size(T_{rsh}) = n\times w_{rsh}\\
  \end{aligned}
\end{equation*}

\begin{equation*}
 \resizebox{0.9\columnwidth}{!}{$
  \begin{aligned}
SizeRatio &= [Size(T_{ust})+Size(T_{idx})+Size(T_{rsh})]/Size(T_{st})\\
&= [w_{idx} + w_{rsh}]/({2^{w_s}\times w_{st}}) + k/n
  \end{aligned}
  $}
\end{equation*}
where $w_{idx}$ and $w_{rsh}$ are the bit width of the values in $T_{idx}$ and $T_{rsh}$, respectively. In our method, however, we force $w_{rsh}$ to be 2, which means that during the self-similarity search process, we limit the value of $t$ in Eq.~\ref{eq:sm} to the range of [0,3]. The value of $w_{idx}$ depends on the number of unique sub-tables and is equal to $floor(log2(k-1))+1$.

\subsection{Higher-Bit Compression}\label{sec:higherBitCompIdea}
Using decomposition and self-similarities can potentially reduce a table's size, especially if the values of a table change continuously. That is, these two compression techniques can be more efficient if there are small differences between consecutive values in a table. On the other hand, there are two issues in the compression of tables with more discrete values that show large differences between consecutive elements.

The first issue is the increase of $w_{st}$ in $T_{st}$ after decomposition (Section~\ref{sec:methodTableDecomp}), which negatively impacts the final table-size savings. This is because there are larger differences between consecutive values in $T$, and therefore the local variations are higher. As a result, the values in $T_{st}$, which stores the local variations, require a longer bit width $w_{st}$. The second issue is with self-similarities (Section~\ref{sec:selfSimilarityIdea}). Since the values of sub-tables are larger, it is likely harder to find similarities among them. Therefore, the number of unique sub-tables increases, which in turn results in lower table-size savings.

As a solution to mitigate these issues, we can split the values of $T$ into higher and lower bits before performing decomposition and self-similarity measures. The values of $T$ are divided into $w_{l}$ lower bits and $w_{out}-w_l$ higher bits, which can be stored in two separate tables $T_{lb}$ and $T_{hb}$, respectively. The table $T_{lb}$  undergoes no compression, but $T_{hb}$ is compressed by using decomposition (Section~\ref{sec:methodTableDecomp}) and self-similarities (Section~\ref{sec:selfSimilarityIdea}).

The intuition behind this practice is to reduce the distances between consecutive values of $T$ by considering higher bits. If we plot both $T$ and $T_{hb}$, the overall shapes of the plots will be similar, however, the slopes of sub-regions in the plot of $T_{hb}$ would be more gentle. Therefore, local variations become lower, which potentially results in more table-size savings after using decomposition and self-similarity techniques.

\subsection{Multilevel Compression}\label{sec:multilevelCompIdea}
Using the three techniques discussed in Sections~\ref{sec:methodTableDecomp}, \ref{sec:selfSimilarityIdea}, and \ref{sec:higherBitCompIdea}, a table $T$ can be significantly compressed and replaced by $T_{lb}$, $T_{ust}$, $T_{idx}$, $T_{rsh}$, and $T_{bias}$. Among these tables, $T_{bias}$ can be compressed further by performing all three techniques on it. As a result, $T_{bias}$ itself is replaced by another set of $T_{lb}$, $T_{ust}$, $T_{idx}$, $T_{rsh}$, and $T_{bias}$. This can potentially achieve further table-size savings in total.

It is worth noting that if we plot the values of $T$ and $T_{bias}$, they will have a similar shape. This is because $T_{bias}$ is the same as $T$ sampled by a factor of $2^{w_{in}-w_s}$. Although $T_{bias}$ has a coarser granularity than $T$, this issue can be resolved by splitting the values of $T_{bias}$ into higher and lower bits, as discussed in Section~\ref{sec:higherBitCompIdea}.

Using the idea of multilevel compression often results in more table-size savings. However, it might increase hardware costs due to the nested decoders needed to retrieve values.

\subsection{Overall Architecture} \label{sec:overallArch}
\begin{figure*} [t!] 
	\centering
		\includegraphics[scale=.5] {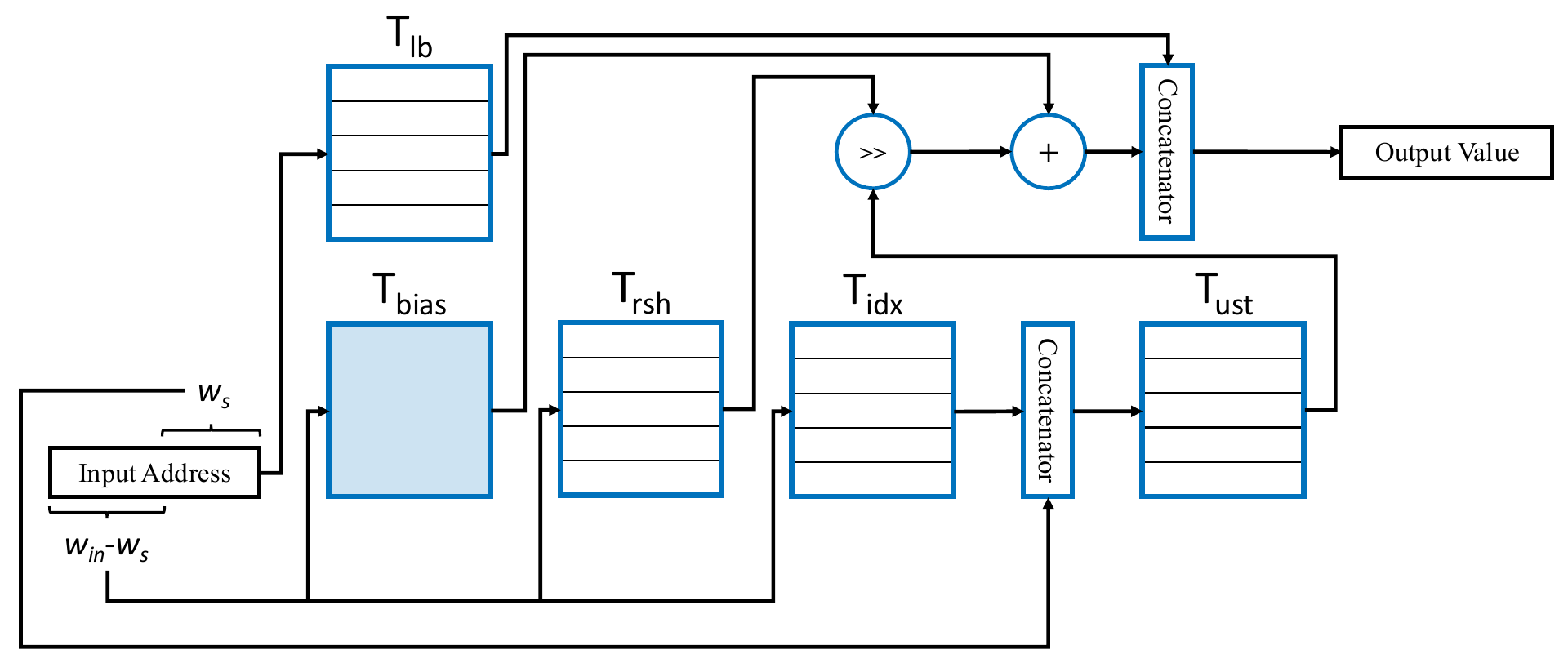}
	\caption{Overall architecture of our CompressedLUT method. In the case of multilevel compression, the same architecture is embedded in T\textsubscript{bias}.}
 \label{fig:overall}
\end{figure*}

Algorithm~\ref{alg:compressedlut} describes the compression techniques used by our CompressedLUT method. This algorithm takes a table $T$ and two parameters $w_s$ and $w_l$ as inputs, and it returns five tables $T_{lb}$, $T_{ust}$, $T_{bias}$, $T_{idx}$, and $T_{rsh}$ as outputs. For multilevel compression, the algorithm must be run again multiple times, given $T_{bias}$ as input. Fig.~\ref{fig:overall} shows the overall architecture of our method.

The parameters $w_s$ and $w_l$ should be determined for each specific input table $T$. In our method, we run the algorithm for different values of the parameters and evaluate them based on the total sizes of all generated tables. Although the runtime of this procedure highly depends on the initial size of a lookup table, our CompressedLUT tool takes around 1.38 seconds on a regular computer to compress a lookup table of 4096 values at 12-bit resolution.

\RestyleAlgo{ruled}
\begin{algorithm}
\caption{CompressedLUT}
\label{alg:compressedlut}
\textbf{Input:} {$T, w_l, w_s$}\\
\textbf{Outputs:} {$T_{lb}, T_{ust}, T_{bias}, T_{idx}, T_{rsh}$} \\ 
\vspace{0.3cm}
$T_{lb}[:] \gets bitand(T[:], 2^{w_{lb}}-1)$\\
$T_{hb}[:] \gets rsh(T[:], w_{lb})$\\

$w_{in} \gets bitwidth(length(T)-1)$\\
$w_{out} \gets bitwidth(max(T))$\\
$n \gets 2^{w_{in}-w_{s}}$\\

\vspace{0.35cm}
\# Compression of $T_{hb}$ Using Decomposition
\vspace{0.15cm}

 \For{\texttt{$i = 1$ to $n$}}{ 
 \texttt{$ST[:] \gets T_{hb}[(i-1) \times 2^{w_s}+1 : i \times 2^{w_s}]$ }\\
    \texttt{$T_{st}[(i-1) \times 2^{w_s}+1 : i \times 2^{w_s}] \gets ST[:]-min(ST[:])$ }\\
    \texttt{$T_{bias}[i] \gets min(ST[:])$}\\
}
\vspace{0.35cm}
  \# Compression of $T_{st}$ Using Self-Similarities\\
  \vspace{0.15cm}
  $SimilarityMatrix[:][:] \gets zeros(n,n)$\\
  $RightShiftMatrix[:][:] \gets zeros(n,n)$\\
 \For{\texttt{$i = 1$ to $n$}}{ 
 $ST_i \gets T_{st}[(i-1) \times 2^{w_s}+1 : i \times 2^{w_s}]$\\
  \For{\texttt{$j = 1$ to $n$}}{ 
  $ST_j \gets T_{st}[(j-1) \times 2^{w_s}+1 : j \times 2^{w_s}]$\\
  \For{\texttt{$t = 0$ to $3$}}{ 
  \If{$rsh(ST_i[:], t) == ST_j[:]$}{
  $SimilarityMatrix[i][j] \gets 1$\\
  $RightShiftMatrix[i][j] \gets t$\\
  $break$
  }
  }}
  }
$k \gets 0$\\
  $SimilarityVector[:] \gets zeros(1,n)$\\
    
\While{$SimilarityMatrix[:][:] ~!= zeros(n,n)$}{  
$k \gets k+1$  \# increment the number of unique sub-tables\\
$SimilarityVector[:] \gets \sum_{i}SimilarityMatrix[i][:]$\\
$idx \gets \arg \max_{i} SimilarityVector[i]$\\
$UST[:] \gets T_{st}[(idx-1) \times 2^{w_s}+1 : idx \times 2^{w_s}]$\\
$T_{ust}[(k-1) \times 2^{w_s}+1 : k \times 2^{w_s}] \gets UST[:]$\\
$T_{idx}[idx] \gets k$\\
$SimilarityMatrix[idx][idx] \gets 0$\\
\# reference similar sub-tables to the $k$th unique sub-table\\
 \For{\texttt{$i = 1$ to $n$}}{ 
 \If{$SimilarityMatrix[i][idx] == 1$}{
 $T_{idx}[i] \gets k$\\
 $T_{rsh} \gets RightShiftMatrix[i][idx]$ \\
$SimilarityMatrix[i][:] \gets zeros(1, n)$\\
$SimilarityMatrix[:][i] \gets zeros(n, 1)$\\
 }
 }
$SimilarityMatrix[idx][:] \gets zeros(1, n)$\\
}
\end{algorithm}

\section{Tool Flow}  \label{sec:toolflow}
\begin{figure} [h] 
	\centering
		\includegraphics[scale=.55] {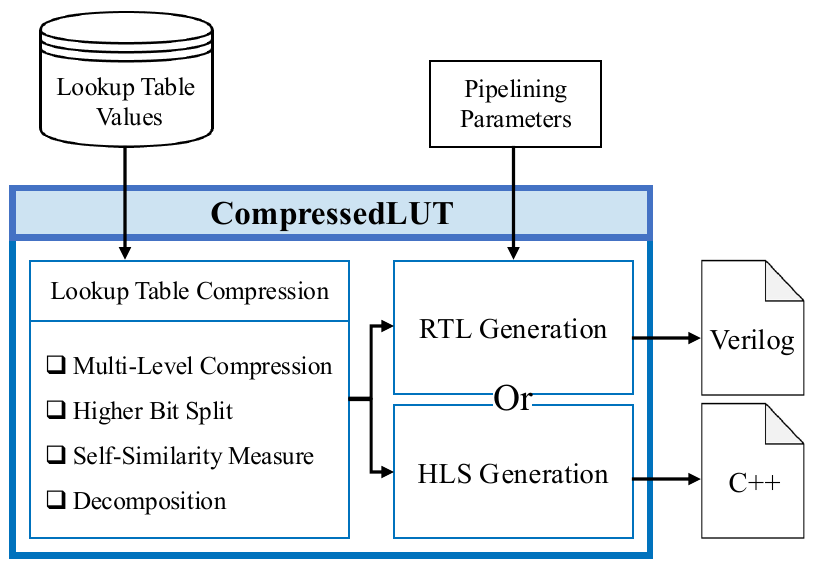}
	\caption{\textcolor{black}{The overview of the CompressedLUT tool flow.}}
 \label{fig:toolflow}
\end{figure}

\textcolor{black}{We developed the CompressedLUT tool to automate the implementation of large lookup tables in either Verilog for RTL or C++ for HLS. The user provides the raw lookup-table values in hexadecimal format, and the tool generates the corresponding RTL or HLS files according to the selected implementation flow. By default, the tool applies all compression techniques described in Sec.~\ref{sec:compressedlut} to maximize table-size reduction. However, users can selectively disable individual techniques to evaluate the trade-off between storage reduction and hardware performance. Fig.~\ref{fig:toolflow} illustrates the CompressedLUT tool flow.}

\begin{figure} [] 
	\centering
		\includegraphics[scale=.43] {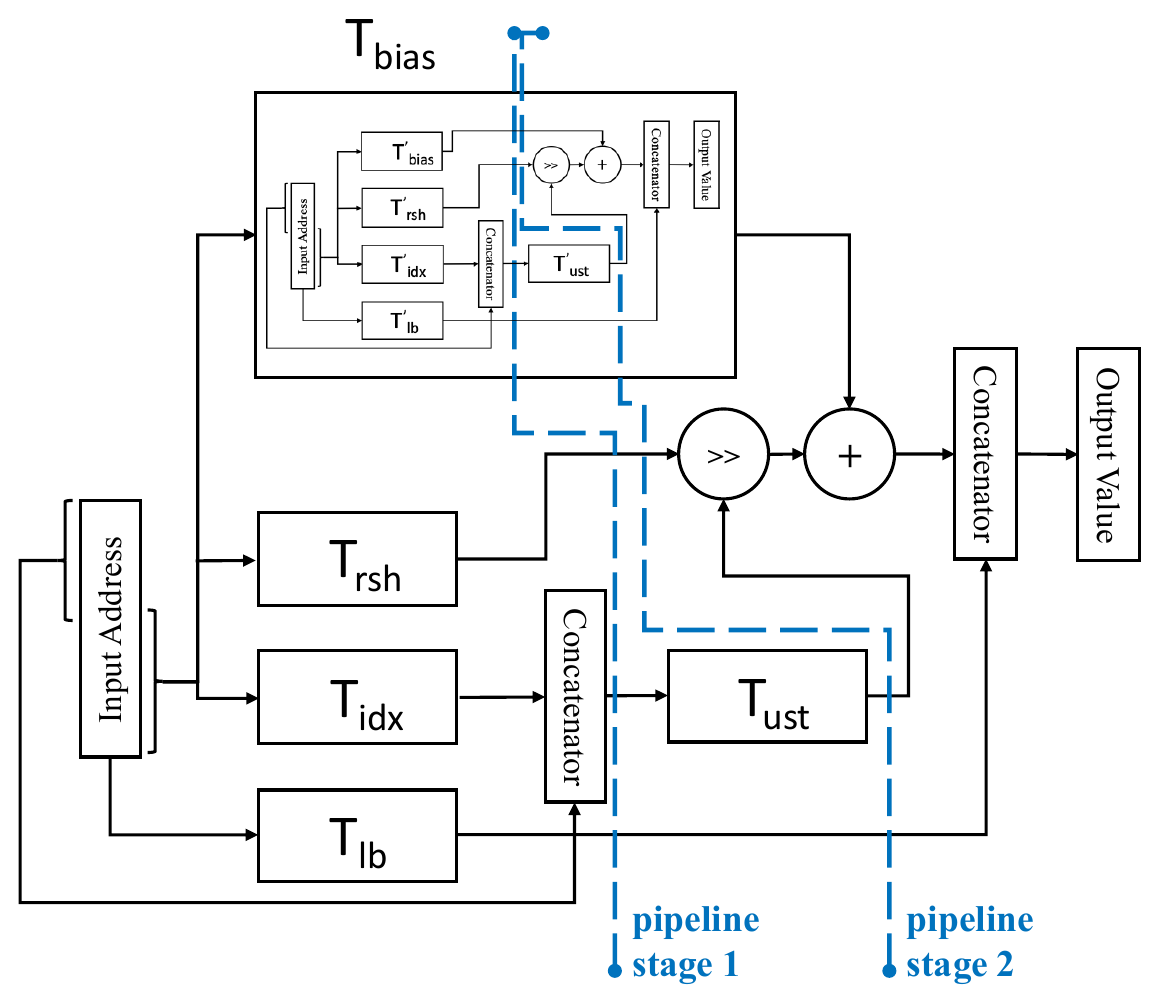}
	\caption{\textcolor{black}{Pipeline register locations in the RTL mode of the CompressedLUT tool.}}
 \label{fig:pipeline}
\end{figure}

\textcolor{black}{In the HLS mode, the tool uses pipelining pragmas to automatically pipeline the architecture and increase throughput. In the RTL mode, the architecture is pipelined by using two predefined pipeline stages, as illustrated in Fig.~\ref{fig:pipeline}. The user can configure the tool to insert pipeline registers at either or both of these locations.}

\section{Implementation Results}
\label{sec:impleresults}
\begin{figure*} [t!] 
	\centering
		\includegraphics[scale=.8] {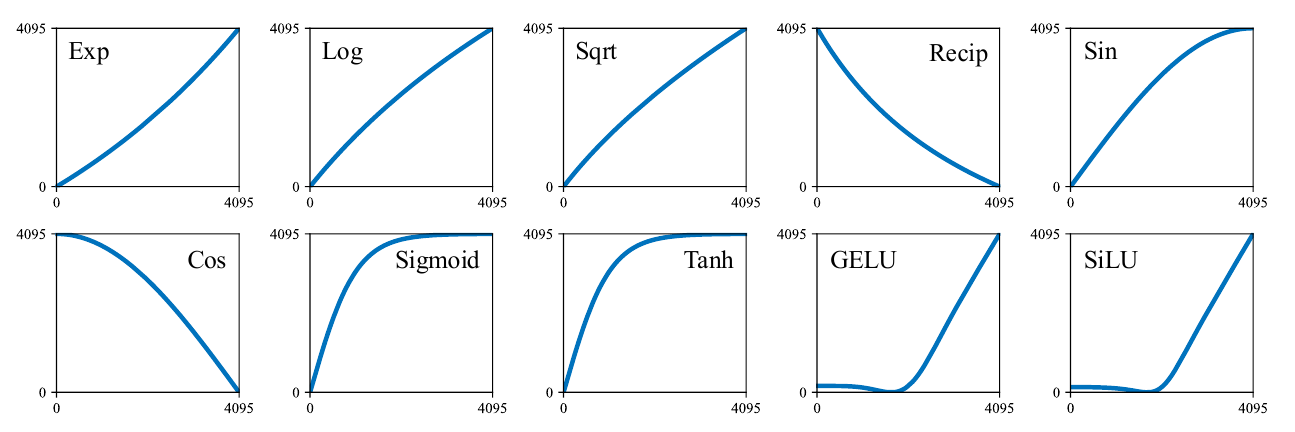}
	\caption{\textcolor{black}{Plots of the implemented nonlinear functions after tabulation. X-axes represent the input address of the tables, and Y-axes represent the corresponding output values.}}
 \label{fig:plots}
\end{figure*} 

\textcolor{black}{Lookup tables are widely used in hardware for various applications, ranging from elementary mathematical functions to novel table-based machine learning models. In this section, we show how CompressedLUT can benefit some of these applications and also compare our method against other approaches, including PlainTable (uncompressed plain table) and TwoTable (two-table decomposition)~\cite{6998028}.}  

\textcolor{black}{In the first part of this section, we implement multiple nonlinear functions at 12-bit resolution, which are frequently used in machine learning and signal processing applications. In the next part, we implement multiple CCMs at 12-bit resolution with different constants. In the last part, we implement two table-based KAN models at 12-bit resolution, which are trained on MNIST and JSC-OpenML datasets.}

\textcolor{black}{Our CompressedLUT tool supports generating hardware designs in either RTL or HLS. In the earlier version of this work~\cite{10.1145/3626202.3637575}, we reported the HLS implementation results. However, in this extended version, we use the tool in its RTL mode. All designs were synthesized and placed and routed using Vivado 2025.2 targeting the xcvu9p-flgb2104-2-i FPGA device, with the Flow\_PerfOptimized\_high setting in the Out-of-Context (OOC) synthesis mode.}

\textcolor{black}{We obtained hardware utilization and timing reports after place and route. We disabled BRAM usage, as the use of BRAMs cannot show the efficiency of compression methods due to the discrete sizes of BRAMs~\cite{9628172}. Additionally, our designs do not utilize any DSP blocks. Therefore, the LUT count served as our primary metric for area measurement. For timing measurement, we implemented a two-stage pipeline for CompressedLUT and a single-stage pipeline for TwoTable with an initiation interval (II) of 1 to achieve maximum throughput. Finally, we adopted Throughput per LUT (TPL) as the comprehensive metric to evaluate and compare the hardware efficiency of the different methods.}

\subsection{Nonlinear Functions}\label{sec:lowresolutionResults}

\begin{table}[]
\caption{\textcolor{black}{Specifications of the implemented nonlinear functions.}}
    \centering
\begin{adjustbox}{width=0.48\textwidth}

\renewcommand{\arraystretch}{1.8}

\begin{tabular}{llc}
\hline
\multicolumn{1}{|c|}{\textbf{Name}} & \multicolumn{1}{c|}{\textbf{Equation}} & \multicolumn{1}{c|}{\textbf{Range}} \\ \hline \hline
\rowcolor[HTML]{EFEFEF} 
\multicolumn{1}{|l|}{\cellcolor[HTML]{EFEFEF}Exp} & \multicolumn{1}{l|}{\cellcolor[HTML]{EFEFEF}$\exp(x)$} & \multicolumn{1}{l|}{\cellcolor[HTML]{EFEFEF}{$x\in[0, \ln(2)]$}} \\ \hline
\multicolumn{1}{|l|}{Log} & \multicolumn{1}{l|}{$\log_2(x)$} & \multicolumn{1}{l|}{{$x\in[1, 2]$}} \\ \hline
\rowcolor[HTML]{EFEFEF} 
\multicolumn{1}{|l|}{\cellcolor[HTML]{EFEFEF}Sqrt} & \multicolumn{1}{l|}{\cellcolor[HTML]{EFEFEF}$\sqrt{x}$} & \multicolumn{1}{l|}{\cellcolor[HTML]{EFEFEF}{$x\in[1, 4]$}} \\ \hline
\multicolumn{1}{|l|}{Recip} & \multicolumn{1}{l|}{$1 / x$} & \multicolumn{1}{l|}{{$x\in[1, 2]$}} \\ \hline
\rowcolor[HTML]{EFEFEF} 
\multicolumn{1}{|l|}{\cellcolor[HTML]{EFEFEF}Sin} & \multicolumn{1}{l|}{\cellcolor[HTML]{EFEFEF}$\sin(x)$} & \multicolumn{1}{l|}{\cellcolor[HTML]{EFEFEF}{$x\in[0, \pi/2]$}} \\ \hline
\multicolumn{1}{|l|}{Cos} & \multicolumn{1}{l|}{$\cos(x)$} & \multicolumn{1}{l|}{{$x\in[0, \pi/2]$}} \\ \hline
\rowcolor[HTML]{EFEFEF} 
\multicolumn{1}{|l|}{\cellcolor[HTML]{EFEFEF}Sigmoid} & \multicolumn{1}{l|}{\cellcolor[HTML]{EFEFEF}$1 / (1 + \exp(-x))$} & \multicolumn{1}{l|}{\cellcolor[HTML]{EFEFEF}{$x\in[0, 8]$}} \\ \hline
\multicolumn{1}{|l|}{Tanh} & \multicolumn{1}{l|}{$\tanh(x)$} & \multicolumn{1}{l|}{{$x\in[0, 4]$}} \\ \hline
\rowcolor[HTML]{EFEFEF} 
\multicolumn{1}{|l|}{\cellcolor[HTML]{EFEFEF}GELU} & \multicolumn{1}{l|}{\cellcolor[HTML]{EFEFEF}$0.5 x \left(1 + \tanh\!\left(\sqrt{2/\pi} \left(x + 0.044715 x^3\right)\right)\right)$} & \multicolumn{1}{l|}{\cellcolor[HTML]{EFEFEF}{$x\in[-4, 4]$}} \\ \hline
\multicolumn{1}{|l|}{SiLU} & \multicolumn{1}{l|}{$x / (1 + \exp(-x))$} & \multicolumn{1}{l|}{{$x\in[-8, 8]$}} \\ \hline
\end{tabular}
\end{adjustbox}
\label{tbl:equations}
\end{table}

\begin{table*}[]
\caption{\textcolor{black}{FPGA implementation results of the nonlinear functions after place~\&~route. Decoder costs are included. TPL (MS/s/LUT) denotes throughput per LUT, measured in megasamples per second per LUT.}}
    \centering
\begin{adjustbox}{width=\textwidth}
\renewcommand{\arraystretch}{1.6}
\begin{tabular}{lclcc|cccccc}
\hline
\multicolumn{3}{|c|}{\textbf{Specifications}} & \multicolumn{2}{c|}{\textbf{Compression}} & \multicolumn{6}{c|}{\textbf{Hardware Costs}} \\ \hline
\multicolumn{1}{|c|}{\textbf{Function}} & \multicolumn{1}{c|}{\textbf{Initial Size} (bit)} & \multicolumn{1}{c|}{\textbf{Method}} & \multicolumn{1}{c|}{\textbf{Final Size} (bit)} & \textbf{Ratio} & \multicolumn{1}{c|}{\textbf{LUT}} & \multicolumn{1}{c|}{\textbf{FF}} & \multicolumn{1}{c|}{\textbf{Fmax} (MHz)} & \multicolumn{1}{c|}{\textbf{Latency} (cy)} & \multicolumn{1}{c|}{\textbf{TPL} (MS/s/LUT)} & \multicolumn{1}{c|}{\textbf{Ratio}} \\ \hline
 \hline
\multicolumn{1}{|l|}{} & \multicolumn{1}{c|}{} & \multicolumn{1}{l|}{\cellcolor[HTML]{EFEFEF}PlainTable} & \cellcolor[HTML]{EFEFEF}49152 & \cellcolor[HTML]{EFEFEF}1.00 & \cellcolor[HTML]{EFEFEF}500 & \cellcolor[HTML]{EFEFEF}37 & \cellcolor[HTML]{EFEFEF}638 & \cellcolor[HTML]{EFEFEF}2 & \cellcolor[HTML]{EFEFEF}1.28 & \multicolumn{1}{c|}{\cellcolor[HTML]{EFEFEF}1.00} \\
\multicolumn{1}{|l|}{} & \multicolumn{1}{c|}{} & \multicolumn{1}{l|}{TwoTable} & 22528 & 0.46 & 218 & 40 & 666 & 3 & 3.06 & \multicolumn{1}{c|}{2.39} \\
\multicolumn{1}{|l|}{\multirow{-3}{*}{Exp}} & \multicolumn{1}{c|}{\multirow{-3}{*}{49152}} & \multicolumn{1}{l|}{\cellcolor[HTML]{EFEFEF}CompressedLUT} & \cellcolor[HTML]{EFEFEF}7836 & \cellcolor[HTML]{EFEFEF}0.16 & \cellcolor[HTML]{EFEFEF}150 & \cellcolor[HTML]{EFEFEF}77 & \cellcolor[HTML]{EFEFEF}781 & \cellcolor[HTML]{EFEFEF}4 & \cellcolor[HTML]{EFEFEF}5.21 & \multicolumn{1}{c|}{\cellcolor[HTML]{EFEFEF}4.08} \\ \hline
\multicolumn{1}{|l|}{} & \multicolumn{1}{c|}{} & \multicolumn{1}{l|}{PlainTable} & 49152 & 1.00 & 508 & 59 & 649 & 2 & 1.28 & \multicolumn{1}{c|}{1.00} \\
\multicolumn{1}{|l|}{} & \multicolumn{1}{c|}{} & \multicolumn{1}{l|}{\cellcolor[HTML]{EFEFEF}TwoTable} & \cellcolor[HTML]{EFEFEF}22528 & \cellcolor[HTML]{EFEFEF}0.46 & \cellcolor[HTML]{EFEFEF}228 & \cellcolor[HTML]{EFEFEF}40 & \cellcolor[HTML]{EFEFEF}644 & \cellcolor[HTML]{EFEFEF}3 & \cellcolor[HTML]{EFEFEF}2.82 & \multicolumn{1}{c|}{\cellcolor[HTML]{EFEFEF}2.21} \\
\multicolumn{1}{|l|}{\multirow{-3}{*}{Log}} & \multicolumn{1}{c|}{\multirow{-3}{*}{49152}} & \multicolumn{1}{l|}{CompressedLUT} & 7314 & 0.15 & 157 & 94 & 796 & 4 & 5.07 & \multicolumn{1}{c|}{3.97} \\ \hline
\multicolumn{1}{|l|}{} & \multicolumn{1}{c|}{} & \multicolumn{1}{l|}{\cellcolor[HTML]{EFEFEF}PlainTable} & \cellcolor[HTML]{EFEFEF}49152 & \cellcolor[HTML]{EFEFEF}1.00 & \cellcolor[HTML]{EFEFEF}482 & \cellcolor[HTML]{EFEFEF}55 & \cellcolor[HTML]{EFEFEF}605 & \cellcolor[HTML]{EFEFEF}2 & \cellcolor[HTML]{EFEFEF}1.26 & \multicolumn{1}{c|}{\cellcolor[HTML]{EFEFEF}1.00} \\
\multicolumn{1}{|l|}{} & \multicolumn{1}{c|}{} & \multicolumn{1}{l|}{TwoTable} & 22528 & 0.46 & 212 & 40 & 664 & 3 & 3.13 & \multicolumn{1}{c|}{2.50} \\
\multicolumn{1}{|l|}{\multirow{-3}{*}{Sqrt}} & \multicolumn{1}{c|}{\multirow{-3}{*}{49152}} & \multicolumn{1}{l|}{\cellcolor[HTML]{EFEFEF}CompressedLUT} & \cellcolor[HTML]{EFEFEF}7940 & \cellcolor[HTML]{EFEFEF}0.16 & \cellcolor[HTML]{EFEFEF}144 & \cellcolor[HTML]{EFEFEF}77 & \cellcolor[HTML]{EFEFEF}801 & \cellcolor[HTML]{EFEFEF}4 & \cellcolor[HTML]{EFEFEF}5.56 & \multicolumn{1}{c|}{\cellcolor[HTML]{EFEFEF}4.43} \\ \hline
\multicolumn{1}{|l|}{} & \multicolumn{1}{c|}{} & \multicolumn{1}{l|}{PlainTable} & 49152 & 1.00 & 508 & 46 & 617 & 2 & 1.21 & \multicolumn{1}{c|}{1.00} \\
\multicolumn{1}{|l|}{} & \multicolumn{1}{c|}{} & \multicolumn{1}{l|}{\cellcolor[HTML]{EFEFEF}TwoTable} & \cellcolor[HTML]{EFEFEF}22528 & \cellcolor[HTML]{EFEFEF}0.46 & \cellcolor[HTML]{EFEFEF}239 & \cellcolor[HTML]{EFEFEF}40 & \cellcolor[HTML]{EFEFEF}657 & \cellcolor[HTML]{EFEFEF}3 & \cellcolor[HTML]{EFEFEF}2.75 & \multicolumn{1}{c|}{\cellcolor[HTML]{EFEFEF}2.26} \\
\multicolumn{1}{|l|}{\multirow{-3}{*}{Recip}} & \multicolumn{1}{c|}{\multirow{-3}{*}{49152}} & \multicolumn{1}{l|}{CompressedLUT} & 9120 & 0.19 & 164 & 81 & 806 & 4 & 4.91 & \multicolumn{1}{c|}{4.05} \\ \hline
\multicolumn{1}{|l|}{} & \multicolumn{1}{c|}{} & \multicolumn{1}{l|}{\cellcolor[HTML]{EFEFEF}PlainTable} & \cellcolor[HTML]{EFEFEF}49152 & \cellcolor[HTML]{EFEFEF}1.00 & \cellcolor[HTML]{EFEFEF}498 & \cellcolor[HTML]{EFEFEF}39 & \cellcolor[HTML]{EFEFEF}657 & \cellcolor[HTML]{EFEFEF}2 & \cellcolor[HTML]{EFEFEF}1.32 & \multicolumn{1}{c|}{\cellcolor[HTML]{EFEFEF}1.00} \\
\multicolumn{1}{|l|}{} & \multicolumn{1}{c|}{} & \multicolumn{1}{l|}{TwoTable} & 22528 & 0.46 & 238 & 40 & 667 & 3 & 2.80 & \multicolumn{1}{c|}{2.12} \\
\multicolumn{1}{|l|}{\multirow{-3}{*}{Sin}} & \multicolumn{1}{c|}{\multirow{-3}{*}{49152}} & \multicolumn{1}{l|}{\cellcolor[HTML]{EFEFEF}CompressedLUT} & \cellcolor[HTML]{EFEFEF}9164 & \cellcolor[HTML]{EFEFEF}0.19 & \cellcolor[HTML]{EFEFEF}172 & \cellcolor[HTML]{EFEFEF}81 & \cellcolor[HTML]{EFEFEF}745 & \cellcolor[HTML]{EFEFEF}4 & \cellcolor[HTML]{EFEFEF}4.33 & \multicolumn{1}{c|}{\cellcolor[HTML]{EFEFEF}3.28} \\ \hline
\multicolumn{1}{|l|}{} & \multicolumn{1}{c|}{} & \multicolumn{1}{l|}{PlainTable} & 49152 & 1.00 & 585 & 55 & 645 & 2 & 1.10 & \multicolumn{1}{c|}{1.00} \\
\multicolumn{1}{|l|}{} & \multicolumn{1}{c|}{} & \multicolumn{1}{l|}{\cellcolor[HTML]{EFEFEF}TwoTable} & \cellcolor[HTML]{EFEFEF}22528 & \cellcolor[HTML]{EFEFEF}0.46 & \cellcolor[HTML]{EFEFEF}238 & \cellcolor[HTML]{EFEFEF}40 & \cellcolor[HTML]{EFEFEF}696 & \cellcolor[HTML]{EFEFEF}3 & \cellcolor[HTML]{EFEFEF}2.92 & \multicolumn{1}{c|}{\cellcolor[HTML]{EFEFEF}2.65} \\
\multicolumn{1}{|l|}{\multirow{-3}{*}{Cos}} & \multicolumn{1}{c|}{\multirow{-3}{*}{49152}} & \multicolumn{1}{l|}{CompressedLUT} & 9164 & 0.19 & 172 & 81 & 770 & 4 & 4.48 & \multicolumn{1}{c|}{4.06} \\ \hline
\multicolumn{1}{|l|}{} & \multicolumn{1}{c|}{} & \multicolumn{1}{l|}{\cellcolor[HTML]{EFEFEF}PlainTable} & \cellcolor[HTML]{EFEFEF}49152 & \cellcolor[HTML]{EFEFEF}1.00 & \cellcolor[HTML]{EFEFEF}371 & \cellcolor[HTML]{EFEFEF}34 & \cellcolor[HTML]{EFEFEF}659 & \cellcolor[HTML]{EFEFEF}2 & \cellcolor[HTML]{EFEFEF}1.78 & \multicolumn{1}{c|}{\cellcolor[HTML]{EFEFEF}1.00} \\
\multicolumn{1}{|l|}{} & \multicolumn{1}{c|}{} & \multicolumn{1}{l|}{TwoTable} & 26624 & 0.54 & 242 & 44 & 723 & 3 & 2.99 & \multicolumn{1}{c|}{1.68} \\
\multicolumn{1}{|l|}{\multirow{-3}{*}{Sigmoid}} & \multicolumn{1}{c|}{\multirow{-3}{*}{49152}} & \multicolumn{1}{l|}{\cellcolor[HTML]{EFEFEF}CompressedLUT} & \cellcolor[HTML]{EFEFEF}13992 & \cellcolor[HTML]{EFEFEF}0.28 & \cellcolor[HTML]{EFEFEF}219 & \cellcolor[HTML]{EFEFEF}67 & \cellcolor[HTML]{EFEFEF}819 & \cellcolor[HTML]{EFEFEF}4 & \cellcolor[HTML]{EFEFEF}3.74 & \multicolumn{1}{c|}{\cellcolor[HTML]{EFEFEF}2.11} \\ \hline
\multicolumn{1}{|l|}{} & \multicolumn{1}{c|}{} & \multicolumn{1}{l|}{PlainTable} & 49152 & 1.00 & 371 & 34 & 659 & 2 & 1.78 & \multicolumn{1}{c|}{1.00} \\
\multicolumn{1}{|l|}{} & \multicolumn{1}{c|}{} & \multicolumn{1}{l|}{\cellcolor[HTML]{EFEFEF}TwoTable} & \cellcolor[HTML]{EFEFEF}26624 & \cellcolor[HTML]{EFEFEF}0.54 & \cellcolor[HTML]{EFEFEF}242 & \cellcolor[HTML]{EFEFEF}44 & \cellcolor[HTML]{EFEFEF}723 & \cellcolor[HTML]{EFEFEF}3 & \cellcolor[HTML]{EFEFEF}2.99 & \multicolumn{1}{c|}{\cellcolor[HTML]{EFEFEF}1.68} \\
\multicolumn{1}{|l|}{\multirow{-3}{*}{Tanh}} & \multicolumn{1}{c|}{\multirow{-3}{*}{49152}} & \multicolumn{1}{l|}{CompressedLUT} & 13992 & 0.28 & 219 & 67 & 819 & 4 & 3.74 & \multicolumn{1}{c|}{2.11} \\ \hline
\multicolumn{1}{|l|}{} & \multicolumn{1}{c|}{} & \multicolumn{1}{l|}{\cellcolor[HTML]{EFEFEF}PlainTable} & \cellcolor[HTML]{EFEFEF}49152 & \cellcolor[HTML]{EFEFEF}1.00 & \cellcolor[HTML]{EFEFEF}445 & \cellcolor[HTML]{EFEFEF}37 & \cellcolor[HTML]{EFEFEF}654 & \cellcolor[HTML]{EFEFEF}2 & \cellcolor[HTML]{EFEFEF}1.47 & \multicolumn{1}{c|}{\cellcolor[HTML]{EFEFEF}1.00} \\
\multicolumn{1}{|l|}{} & \multicolumn{1}{c|}{} & \multicolumn{1}{l|}{TwoTable} & 24576 & 0.50 & 222 & 39 & 682 & 3 & 3.07 & \multicolumn{1}{c|}{2.09} \\
\multicolumn{1}{|l|}{\multirow{-3}{*}{GELU}} & \multicolumn{1}{c|}{\multirow{-3}{*}{49152}} & \multicolumn{1}{l|}{\cellcolor[HTML]{EFEFEF}CompressedLUT} & \cellcolor[HTML]{EFEFEF}12640 & \cellcolor[HTML]{EFEFEF}0.26 & \cellcolor[HTML]{EFEFEF}172 & \cellcolor[HTML]{EFEFEF}66 & \cellcolor[HTML]{EFEFEF}760 & \cellcolor[HTML]{EFEFEF}4 & \cellcolor[HTML]{EFEFEF}4.42 & \multicolumn{1}{c|}{\cellcolor[HTML]{EFEFEF}3.01} \\ \hline
\multicolumn{1}{|l|}{} & \multicolumn{1}{c|}{} & \multicolumn{1}{l|}{PlainTable} & 49152 & 1.00 & 415 & 40 & 657 & 2 & 1.58 & \multicolumn{1}{c|}{1.00} \\
\multicolumn{1}{|l|}{} & \multicolumn{1}{c|}{} & \multicolumn{1}{l|}{\cellcolor[HTML]{EFEFEF}TwoTable} & \cellcolor[HTML]{EFEFEF}22528 & \cellcolor[HTML]{EFEFEF}0.46 & \cellcolor[HTML]{EFEFEF}209 & \cellcolor[HTML]{EFEFEF}40 & \cellcolor[HTML]{EFEFEF}705 & \cellcolor[HTML]{EFEFEF}3 & \cellcolor[HTML]{EFEFEF}3.37 & \multicolumn{1}{c|}{\cellcolor[HTML]{EFEFEF}2.13} \\
\multicolumn{1}{|l|}{\multirow{-3}{*}{SiLU}} & \multicolumn{1}{c|}{\multirow{-3}{*}{49152}} & \multicolumn{1}{l|}{CompressedLUT} & 9088 & 0.18 & 175 & 81 & 824 & 4 & 4.71 & \multicolumn{1}{c|}{2.97} \\ \hline
\end{tabular}
\end{adjustbox}
\label{tbl:functions}
\end{table*}

A low-resolution function at up to 12 bits can be directly evaluated by lookup tables containing the function's values for all possible input combinations~\cite{9628172}. Such tables can be compressed using our lossless compression method, which can reduce hardware costs without loss of accuracy.

\textcolor{black}{For this case study, we targeted a number of nonlinear functions at 12-bit resolution, each of which had a baseline table of $12\times2^{12} = 49152$ bits. The equation and input range of each function are provided in Table~\ref{tbl:equations}.} The minimum value of each table is subtracted from all the values in that table, which could potentially remove excessive output bits such as sign bits in some cases. \textcolor{black}{The output values are then scaled to span the range $[0, 1]$ and quantized to 12-bit resolution. Fig.~\ref{fig:plots} shows the plots of implemented nonlinear functions after tabulation.} We used CompressedLUT as well as PlainTable and TwoTable~\cite{6998028} methods to compress the tables as much as possible. As in~\cite{6998028}, we used the total bit count as a metric to guide the selection of decomposition parameters in each method.

\begin{figure} [] 
	\centering
		\includegraphics[scale=.33] {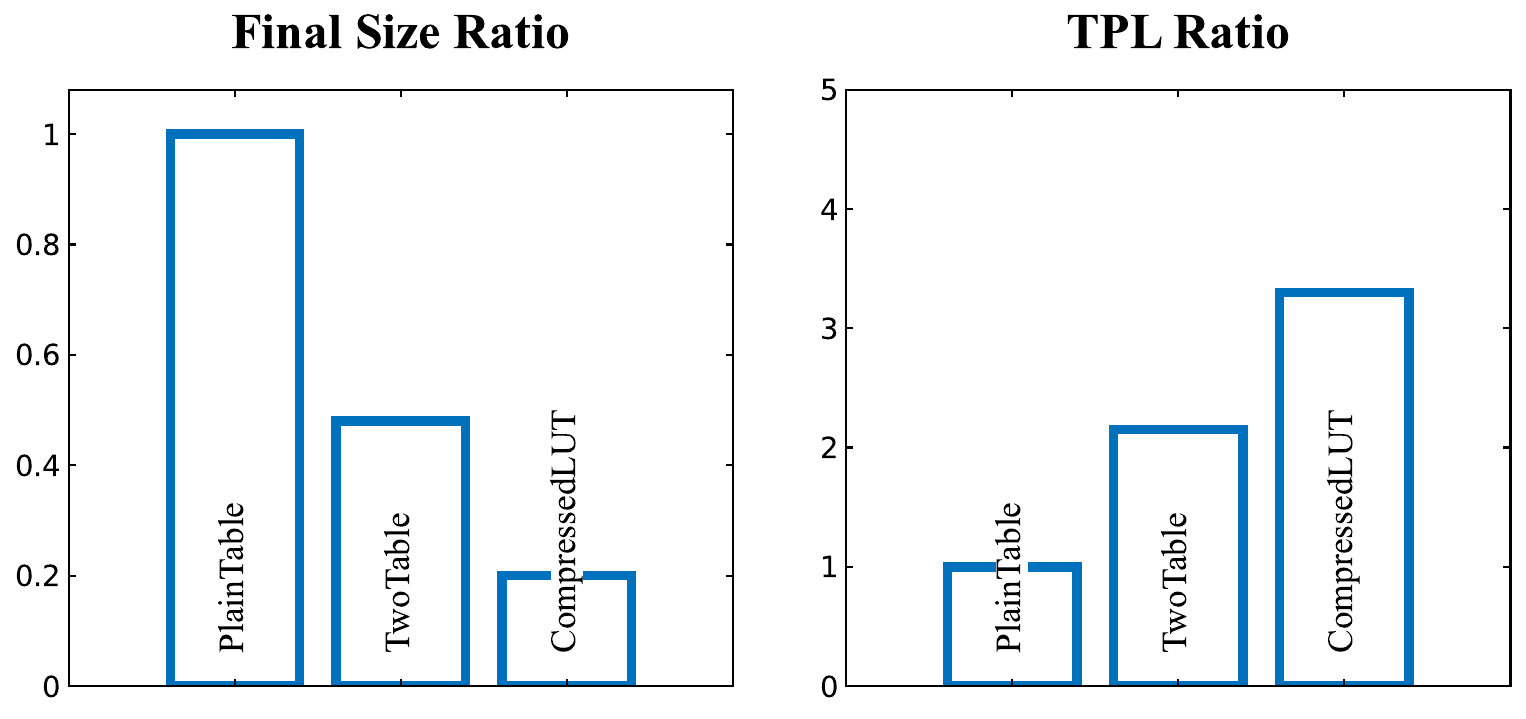}
	\caption{\textcolor{black}{Average FPGA implementation results of the nonlinear functions.}}
\label{fig:functions_avg}
\end{figure}

\textcolor{black}{Table~\ref{tbl:functions} shows the FPGA implementation results of the nonlinear functions. Table~\ref{tbl:functions_avg} and Fig.~\ref{fig:functions_avg} summarize the average results obtained using each implementation method. ``Initial Size" and ``Final Size" show the total bit count before and after compression, respectively. \textcolor{black}{``Fmax" shows the maximum achievable clock frequency in megahertz (MHz), and ``Latency" shows the time required for data to propagate from input to output, measured in the number of clock cycles (cy). Finally, ``TPL" shows throughput per LUT in megasamples per second per LUT (MS/s/LUT).} As seen, our method can compress the tables on average by 80\%, whereas TwoTable can compress them on average by 52\%. In terms of TPL, our method is 3.30 times better than PlainTable, whereas TwoTable is 2.15 times better than PlainTable.}

\begin{table}[]
\caption{\textcolor{black}{Average FPGA implementation results of the nonlinear functions.}}
\centering
\renewcommand{\arraystretch}{1.6}
\setlength{\tabcolsep}{16pt}
\begin{tabular}{|l|c|c|}
\hline
\multicolumn{1}{|c|}{\textbf{Method}} & \textbf{Final Size Ratio} & \textbf{TPL Ratio} \\ \hline \hline
\rowcolor[HTML]{EFEFEF} 
PlainTable & 1.00 & 1.00 \\ \hline
TwoTable & 0.48 & 2.15 \\ \hline
\rowcolor[HTML]{EFEFEF} 
CompressedLUT & 0.20 & 3.30 \\ \hline
\end{tabular}
\label{tbl:functions_avg}
\end{table}

Unlike low-resolution functions, it is not practical to fully tabulate the values of a function beyond 12-bit resolutions due to the exponentially growing size of the resulting tables. In such cases, approximate methods, such as BT, MT, and PPA can be applied to reduce overall table size at the expense of accuracy. As discussed earlier, these approximate methods still rely on lookup tables to store essential values to perform computations. For instance, BT and MT methods rely on TIV and TO tables. In addition, PPA methods store the coefficients of polynomials in lookup tables. Our CompressedLUT method can be plugged into such table-based methods to compress their tables, which reduces hardware costs with no additional approximation error. However, compressing such tables is not as easy as compressing the tables of low-resolution functions. This is because the lookup tables used in table-based methods usually do not show smooth local variations compared to the tables of low-resolution functions. For instance, the TIV table of a function, implemented by an MT method, contains uniformly sampled values of the function. Therefore, the difference between every two consecutive values in the TIV table is likely larger than that of two consecutive values in the plain table of the function. Nonetheless, our method can achieve significant savings in table size due to breaking output values into higher bits and lower bits as well as using a multilevel compression technique.

\subsection{Constant Coefficient Multipliers (CCMs)}

\textcolor{black}{Multiplication is a fundamental operation in many applications, including digital signal processing and machine learning. In applications where one operand has a fixed value, the multiplication can be implemented using a CCM that can be performed using specialized hardware rather than a general-purpose multiplier~\cite{7752883, 5753874, Walters2017ReducedAreaCA, 9114822, 10.1145/3494570, 10323844}.} 

\textcolor{black}{Since a CCM is a univariate linear function, one possible implementation is a lookup table in which the precomputed values of the function are stored for every possible input value. The output can then be obtained directly by using the input as the address of the table. This approach does not introduce any additional approximation error other than input and output quantizations. CompressedLUT can efficiently compress such tables without additional loss of accuracy.}

\textcolor{black}{For this case study, we developed a script to generate lookup tables for CCMs for a variety of constant coefficients. The script evaluated \(y=cx\) over the input domain $[0, 1)$ at 12-bit input and output bitwidths. We then applied CompressedLUT to each generated table to reduce hardware costs. In addition, we implemented CCMs using other methods for comparison, including Vivado, PlainTable, and TwoTable. The Vivado method represents the naive implementation, in which each CCM is described directly in Verilog as \texttt{assign y = c * x}, allowing Vivado to synthesize and implement the constant multiplication on the FPGA. Depending on the constant value, \(c\) was represented using 21--24 bits to provide sufficient internal precision such that the rounded outputs matched the accurate 12-bit reference values generated by our table-generation script.}

\begin{figure} [] 
	\centering
		\includegraphics[scale=.33] {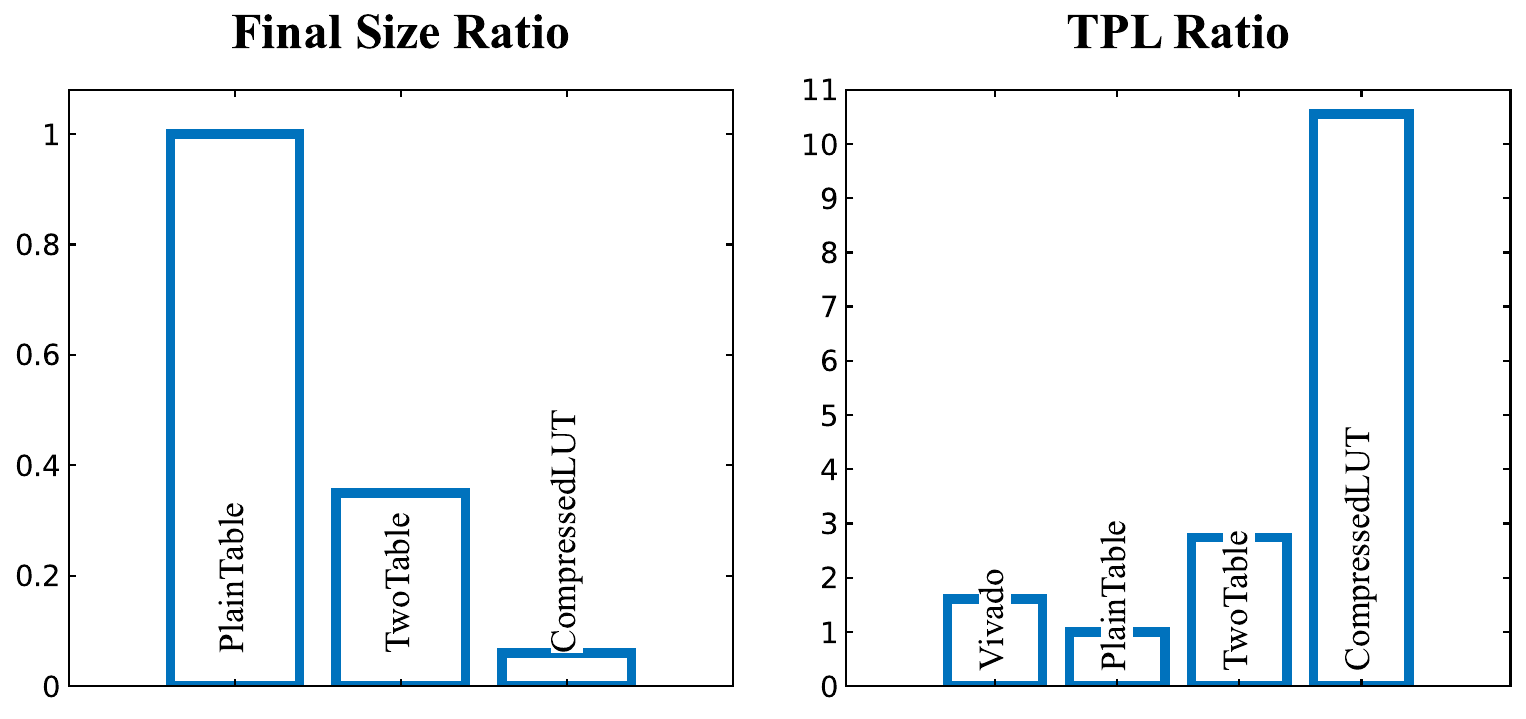}
	\caption{\textcolor{black}{Average FPGA implementation results of the CCMs.}}
 \label{fig:ccm_avg}
\end{figure}

\textcolor{black}{Table~\ref{tbl:ccm} shows the FPGA implementation results of the CCMs. Table~\ref{tbl:ccm_avg} and Fig.~\ref{fig:ccm_avg} summarize the average results obtained using each implementation method. The results indicate that our method can compress the tables on average by 94\%, whereas TwoTable can compress them on average by 65\%. In terms of TPL, our method is 10.55 times better than PlainTable, whereas Vivado and TwoTable are 1.61 and 2.74 times better than PlainTable, respectively.} 

\subsection{Kolmogorov-Arnold Networks (KANs)}

\begin{table*}[]
\caption{\textcolor{black}{FPGA implementation results of the CCMs after place~\&~route. TPL (MS/s/LUT) denotes throughput per LUT, measured in megasamples per second per LUT.}}
    \centering
\setlength{\tabcolsep}{5pt}
\renewcommand{\arraystretch}{1.6}
\begin{tabular}{|ccl|cc|cccccc|}
\hline
\multicolumn{3}{|c|}{\textbf{Specifications}} & \multicolumn{2}{c|}{\textbf{Compression}} & \multicolumn{6}{c|}{\textbf{Hardware Costs}} \\ \hline
\multicolumn{1}{|c|}{\textbf{Constant}} & \multicolumn{1}{c|}{\textbf{Initial Size} (bit)} & \multicolumn{1}{c|}{\textbf{Method}} & \multicolumn{1}{c|}{\textbf{Final Size} (bit)} & \textbf{Ratio} & \multicolumn{1}{c|}{\textbf{LUT}} & \multicolumn{1}{c|}{\textbf{FF}} & \multicolumn{1}{c|}{\textbf{Fmax} (MHz)} & \multicolumn{1}{c|}{\textbf{Latency} (cy)} & \multicolumn{1}{c|}{\textbf{TPL} (MS/s/LUT)} & \textbf{Ratio} \\ \hline \hline
\multicolumn{1}{|c|}{} & \multicolumn{1}{c|}{-} & \cellcolor[HTML]{EFEFEF}Vivado & \cellcolor[HTML]{EFEFEF}- & \cellcolor[HTML]{EFEFEF}- & \cellcolor[HTML]{EFEFEF}133 & \cellcolor[HTML]{EFEFEF}23 & \cellcolor[HTML]{EFEFEF}284 & \cellcolor[HTML]{EFEFEF}2 & \cellcolor[HTML]{EFEFEF}2.14 & \cellcolor[HTML]{EFEFEF}1.37 \\
\multicolumn{1}{|c|}{} & \multicolumn{1}{c|}{} & PlainTable & 49152 & 1.00 & 390 & 32 & 608 & 2 & 1.56 & 1.00 \\
\multicolumn{1}{|c|}{} & \multicolumn{1}{c|}{} & \cellcolor[HTML]{EFEFEF}TwoTable & \cellcolor[HTML]{EFEFEF}13824 & \cellcolor[HTML]{EFEFEF}0.28 & \cellcolor[HTML]{EFEFEF}154 & \cellcolor[HTML]{EFEFEF}36 & \cellcolor[HTML]{EFEFEF}677 & \cellcolor[HTML]{EFEFEF}3 & \cellcolor[HTML]{EFEFEF}4.40 & \cellcolor[HTML]{EFEFEF}2.82 \\
\multicolumn{1}{|c|}{\multirow{-4}{*}{$e^{-1}$}} & \multicolumn{1}{c|}{\multirow{-3}{*}{49152}} & CompressedLUT & 2624 & 0.05 & 47 & 68 & 794 & 4 & 16.89 & 10.84 \\ \hline
\multicolumn{1}{|c|}{} & \multicolumn{1}{c|}{-} & \cellcolor[HTML]{EFEFEF}Vivado & \cellcolor[HTML]{EFEFEF}- & \cellcolor[HTML]{EFEFEF}- & \cellcolor[HTML]{EFEFEF}92 & \cellcolor[HTML]{EFEFEF}24 & \cellcolor[HTML]{EFEFEF}307 & \cellcolor[HTML]{EFEFEF}2 & \cellcolor[HTML]{EFEFEF}3.34 & \cellcolor[HTML]{EFEFEF}2.49 \\
\multicolumn{1}{|c|}{} & \multicolumn{1}{c|}{} & PlainTable & 49152 & 1.00 & 467 & 35 & 625 & 2 & 1.34 & 1.00 \\
\multicolumn{1}{|c|}{} & \multicolumn{1}{c|}{} & \cellcolor[HTML]{EFEFEF}TwoTable & \cellcolor[HTML]{EFEFEF}18432 & \cellcolor[HTML]{EFEFEF}0.38 & \cellcolor[HTML]{EFEFEF}174 & \cellcolor[HTML]{EFEFEF}39 & \cellcolor[HTML]{EFEFEF}745 & \cellcolor[HTML]{EFEFEF}3 & \cellcolor[HTML]{EFEFEF}4.28 & \cellcolor[HTML]{EFEFEF}3.20 \\
\multicolumn{1}{|c|}{\multirow{-4}{*}{$\log(2)$}} & \multicolumn{1}{c|}{\multirow{-3}{*}{49152}} & CompressedLUT & 2976 & 0.06 & 54 & 73 & 864 & 4 & 16.00 & 11.96 \\ \hline
\multicolumn{1}{|c|}{} & \multicolumn{1}{c|}{-} & \cellcolor[HTML]{EFEFEF}Vivado & \cellcolor[HTML]{EFEFEF}- & \cellcolor[HTML]{EFEFEF}- & \cellcolor[HTML]{EFEFEF}150 & \cellcolor[HTML]{EFEFEF}24 & \cellcolor[HTML]{EFEFEF}262 & \cellcolor[HTML]{EFEFEF}2 & \cellcolor[HTML]{EFEFEF}1.75 & \cellcolor[HTML]{EFEFEF}0.91 \\
\multicolumn{1}{|c|}{} & \multicolumn{1}{c|}{} & PlainTable & 49152 & 1.00 & 347 & 30 & 664 & 2 & 1.91 & 1.00 \\
\multicolumn{1}{|c|}{} & \multicolumn{1}{c|}{} & \cellcolor[HTML]{EFEFEF}TwoTable & \cellcolor[HTML]{EFEFEF}18432 & \cellcolor[HTML]{EFEFEF}0.38 & \cellcolor[HTML]{EFEFEF}174 & \cellcolor[HTML]{EFEFEF}39 & \cellcolor[HTML]{EFEFEF}720 & \cellcolor[HTML]{EFEFEF}3 & \cellcolor[HTML]{EFEFEF}4.14 & \cellcolor[HTML]{EFEFEF}2.16 \\
\multicolumn{1}{|c|}{\multirow{-4}{*}{$1/\sqrt{2}$}} & \multicolumn{1}{c|}{\multirow{-3}{*}{49152}} & CompressedLUT & 2864 & 0.06 & 53 & 73 & 824 & 4 & 15.55 & 8.12 \\ \hline
\multicolumn{1}{|c|}{} & \multicolumn{1}{c|}{-} & \cellcolor[HTML]{EFEFEF}Vivado & \cellcolor[HTML]{EFEFEF}- & \cellcolor[HTML]{EFEFEF}- & \cellcolor[HTML]{EFEFEF}114 & \cellcolor[HTML]{EFEFEF}24 & \cellcolor[HTML]{EFEFEF}294 & \cellcolor[HTML]{EFEFEF}2 & \cellcolor[HTML]{EFEFEF}2.58 & \cellcolor[HTML]{EFEFEF}1.98 \\
\multicolumn{1}{|c|}{} & \multicolumn{1}{c|}{} & PlainTable & 49152 & 1.00 & 455 & 33 & 593 & 2 & 1.30 & 1.00 \\
\multicolumn{1}{|c|}{} & \multicolumn{1}{c|}{} & \cellcolor[HTML]{EFEFEF}TwoTable & \cellcolor[HTML]{EFEFEF}18432 & \cellcolor[HTML]{EFEFEF}0.38 & \cellcolor[HTML]{EFEFEF}196 & \cellcolor[HTML]{EFEFEF}39 & \cellcolor[HTML]{EFEFEF}706 & \cellcolor[HTML]{EFEFEF}3 & \cellcolor[HTML]{EFEFEF}3.60 & \cellcolor[HTML]{EFEFEF}2.76 \\
\multicolumn{1}{|c|}{\multirow{-4}{*}{$\pi/4$}} & \multicolumn{1}{c|}{\multirow{-3}{*}{49152}} & CompressedLUT & 2976 & 0.06 & 55 & 73 & 794 & 4 & 14.44 & 11.08 \\ \hline
\multicolumn{1}{|c|}{} & \multicolumn{1}{c|}{-} & \cellcolor[HTML]{EFEFEF}Vivado & \cellcolor[HTML]{EFEFEF}- & \cellcolor[HTML]{EFEFEF}- & \cellcolor[HTML]{EFEFEF}150 & \cellcolor[HTML]{EFEFEF}24 & \cellcolor[HTML]{EFEFEF}356 & \cellcolor[HTML]{EFEFEF}2 & \cellcolor[HTML]{EFEFEF}2.37 & \cellcolor[HTML]{EFEFEF}1.77 \\
\multicolumn{1}{|c|}{} & \multicolumn{1}{c|}{} & PlainTable & 49152 & 1.00 & 479 & 36 & 644 & 2 & 1.34 & 1.00 \\
\multicolumn{1}{|c|}{} & \multicolumn{1}{c|}{} & \cellcolor[HTML]{EFEFEF}TwoTable & \cellcolor[HTML]{EFEFEF}18432 & \cellcolor[HTML]{EFEFEF}0.38 & \cellcolor[HTML]{EFEFEF}185 & \cellcolor[HTML]{EFEFEF}39 & \cellcolor[HTML]{EFEFEF}711 & \cellcolor[HTML]{EFEFEF}3 & \cellcolor[HTML]{EFEFEF}3.84 & \cellcolor[HTML]{EFEFEF}2.86 \\
\multicolumn{1}{|c|}{\multirow{-4}{*}{$\sqrt{3}/2$}} & \multicolumn{1}{c|}{\multirow{-3}{*}{49152}} & CompressedLUT & 2920 & 0.06 & 54 & 73 & 812 & 4 & 15.04 & 11.18 \\ \hline
\end{tabular}
 \label{tbl:ccm}
\end{table*}

\begin{table}[]
\caption{\textcolor{black}{Average FPGA implementation results of the CCMs.}}
\centering
\setlength{\tabcolsep}{16pt}
\renewcommand{\arraystretch}{1.6}
\begin{tabular}{|l|c|c|}
\hline
\multicolumn{1}{|c|}{\textbf{Method}} & \textbf{Final Size Ratio} & \textbf{TPL Ratio} \\ \hline \hline
\rowcolor[HTML]{EFEFEF} 
Vivado & - & 1.61 \\ \hline
PlainTable & 1.00 & 1.00  \\ \hline
\rowcolor[HTML]{EFEFEF} 
TwoTable & 0.35 & 2.74 \\ \hline
CompressedLUT & 0.06 & 10.55 \\ \hline
\end{tabular}
 \label{tbl:ccm_avg}
\end{table}

\begin{figure} [] 
	\centering
		\includegraphics[scale=.34] {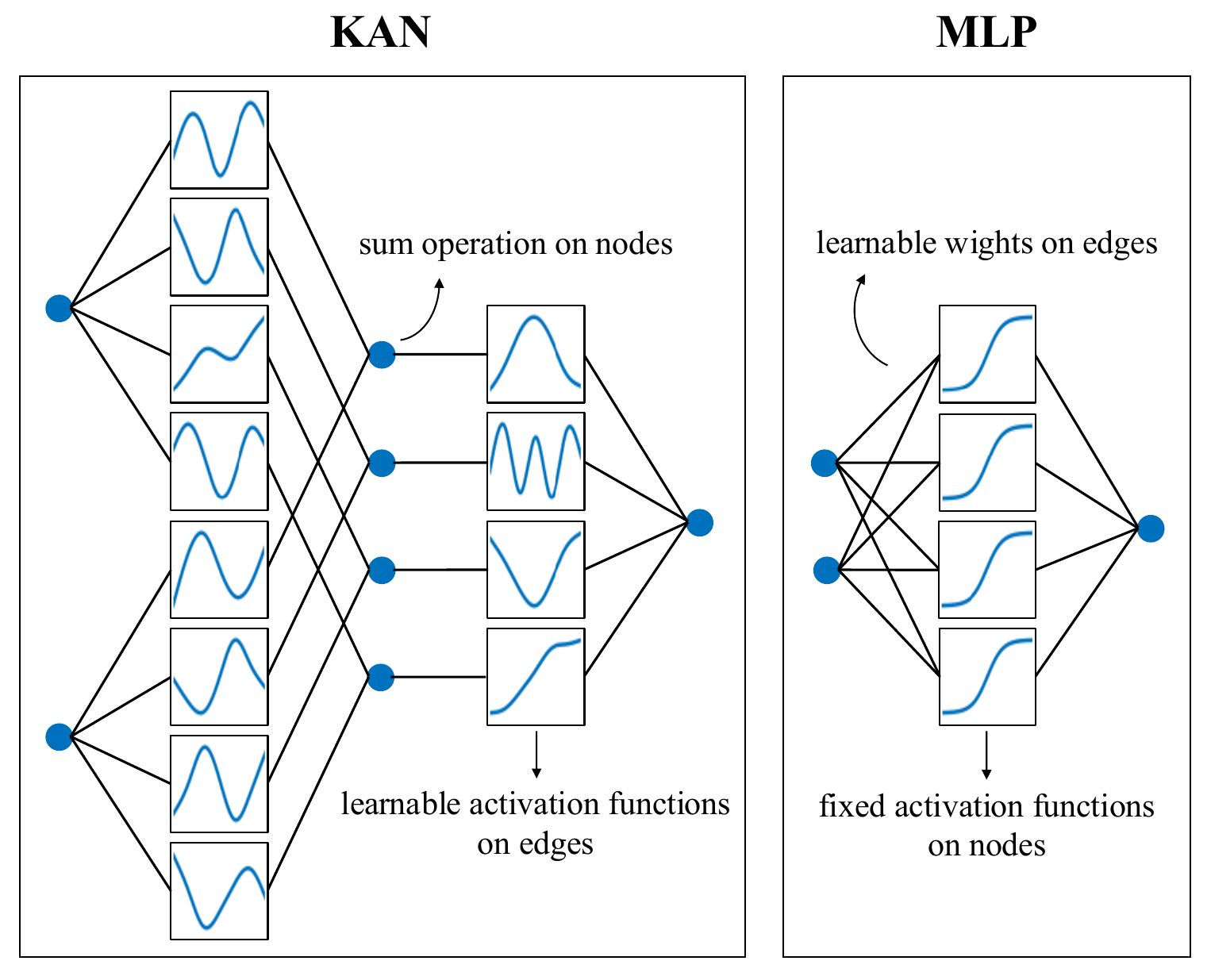}
	\caption{\textcolor{black}{Difference between KAN and MLP architectures~\cite{ICLR2025_afaed896}.}}
 \label{fig:kan_vs_mlp}
\end{figure}

 \textcolor{black}{KANs~\cite{ICLR2025_afaed896} are a promising alternative to MLPs, inspired by the Kolmogorov--Arnold representation theorem. The KAN paper demonstrated that these models can provide improved accuracy and interpretability compared with MLPs, particularly for function fitting problems. Unlike MLPs that perform fixed nonlinear activation functions on nodes, KANs apply learnable nonlinear activation functions on edges. Fig.~\ref{fig:kan_vs_mlp} shows the difference between KAN and MLP architectures.}

\textcolor{black}{KANEL\'E~\cite{10.1145/3748173.3779202} provides a framework for implementing KANs on FPGAs using lookup tables for function evaluation. After quantization-aware training and pruning, KANEL\'E implements the learned edge functions as lookup tables. Therefore, the evaluation of edge functions during inference is replaced by direct table lookups. The table outputs associated with each node are accumulated using balanced, pipelined adder trees. This table-based method avoids the costly evaluation of edge functions and enables low-latency, high-throughput KAN inference on FPGAs.}

\textcolor{black}{Despite these benefits, a table-based KAN may require a large number of tables. Before pruning, a fully connected KAN layer with $d_{\mathrm{in}}$ inputs and $d_{\mathrm{out}}$ outputs contains as many as $d_{\mathrm{in}}d_{\mathrm{out}}$ learnable univariate edge functions, each of which is mapped to a separate lookup table. Furthermore, the number of entries in each table increases exponentially with its input bitwidth. The storage cost can therefore become substantial in high-resolution KAN implementations containing many nonlinear edge functions. This characteristic makes KANEL\'E an appropriate application for CompressedLUT. Our method reduces redundancy in the tables representing the learned edge functions without retraining the network or introducing additional approximation errors.}

\textcolor{black}{For this case study, we used KANEL\'E to train the KAN models on two different classification datasets, including MNIST and JSC-OpenML. Next, CompressedLUT was applied to the tables associated with the KAN layers generated by KANEL\'E to reduce hardware costs. The original KANEL\'E evaluation~\cite{10.1145/3748173.3779202} assigned different, generally lower, quantization bitwidths to individual layers to balance accuracy and hardware costs. In our experiments, all layers were instead quantized to 12-bit resolution. This higher bitwidth produced slightly higher accuracy than the bitwidths used in the original work. Nevertheless, the purpose of this evaluation was not to analyze the relationship between quantization precision and prediction accuracy or to optimize KANEL\'E through hyperparameter tuning. Rather, it was designed as a case study to determine how effectively CompressedLUT reduces the hardware costs of table-based KAN layers at such a high resolution. In practice, using 12-bit precision for every layer may not be necessary, and the appropriate bitwidth should be selected for each layer according to the accuracy and hardware constraints of the target application.}

\begin{figure} [] 
	\centering
		\includegraphics[scale=.33] {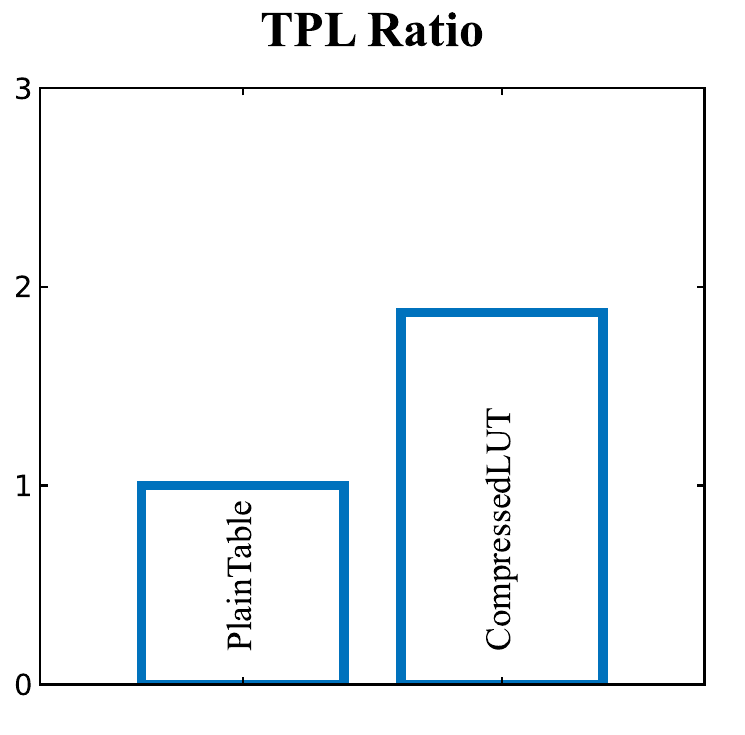}
	\caption{\textcolor{black}{Average FPGA implementation results of the KANs.}}
 \label{fig:kan_avg}
\end{figure}

\textcolor{black}{Table~\ref{tbl:kan} shows the FPGA implementation results of the KANs. Table~\ref{tbl:kan_avg} and Fig.~\ref{fig:kan_avg} summarize the average results obtained using each implementation method. As seen, our method on average has 1.87 times higher TPL than PlainTable.}

\begin{table*}[]
\caption{\textcolor{black}{FPGA implementation results of the KANs after place~\&~route. TPL (kS/s/LUT) denotes throughput per LUT, measured in kilosamples per second per LUT.}}
    \centering
\setlength{\tabcolsep}{9pt}
\renewcommand{\arraystretch}{1.6}
\begin{tabular}{|lcl|cccccc|}
\hline
\multicolumn{3}{|c|}{\textbf{Specifications}} & \multicolumn{6}{c|}{\textbf{Hardware Costs}} \\ \hline \hline
\multicolumn{1}{|c|}{\textbf{Dataset}} & \multicolumn{1}{c|}{\textbf{Accuracy}} & \multicolumn{1}{c|}{\textbf{Method}} & \multicolumn{1}{c|}{\textbf{LUT}} & \multicolumn{1}{c|}{\textbf{FF}} & \multicolumn{1}{c|}{\textbf{Fmax} (MHz)} & \multicolumn{1}{c|}{\textbf{Latency} (cy)} & \multicolumn{1}{c|}{\textbf{TPL} (kS/s/LUT)} & \textbf{Ratio} \\ \hline
\multicolumn{1}{|l|}{} & \multicolumn{1}{c|}{} & \cellcolor[HTML]{EFEFEF}PlainTable & \cellcolor[HTML]{EFEFEF}351816 & \cellcolor[HTML]{EFEFEF}174128 & \cellcolor[HTML]{EFEFEF}182 & \cellcolor[HTML]{EFEFEF}10 & \cellcolor[HTML]{EFEFEF}0.52 & \cellcolor[HTML]{EFEFEF}1.00 \\
\multicolumn{1}{|l|}{\multirow{-2}{*}{MNIST}} & \multicolumn{1}{c|}{\multirow{-2}{*}{97.08\%}} & CompressedLUT & 266064 & 194362 & 291 & 11 & 1.09 & 2.11 \\ \hline
\multicolumn{1}{|l|}{} & \multicolumn{1}{c|}{} & \cellcolor[HTML]{EFEFEF}PlainTable & \cellcolor[HTML]{EFEFEF}63592 & \cellcolor[HTML]{EFEFEF}2922 & \cellcolor[HTML]{EFEFEF}516 & \cellcolor[HTML]{EFEFEF}7 & \cellcolor[HTML]{EFEFEF}8.11 & \cellcolor[HTML]{EFEFEF}1.00 \\
\multicolumn{1}{|l|}{\multirow{-2}{*}{JSC-OpenML}} & \multicolumn{1}{c|}{\multirow{-2}{*}{76.34\%}} & CompressedLUT & 30858 & 8088 & 414 & 9 & 13.42 & 1.65 \\ \hline
\end{tabular}
 \label{tbl:kan}
\end{table*}

\begin{table}[]
\caption{\textcolor{black}{Average FPGA implementation results of the KANs.}}
\centering
\renewcommand{\arraystretch}{1.6}
\setlength{\tabcolsep}{37pt}
\begin{tabular}{|l|c|c|}
\hline
\multicolumn{1}{|c|}{\textbf{Method}}  & \textbf{TPL Ratio} \\ \hline \hline
\rowcolor[HTML]{EFEFEF} 
PlainTable &  1.00 \\ \hline
CompressedLUT &  1.87 \\ \hline

\end{tabular}
 \label{tbl:kan_avg}
\end{table}

\section{Conclusions}
\label{sec:conclusions}
\textcolor{black}{In this paper, we proposed CompressedLUT, a lossless lookup table compression method that combines multilevel compression, decomposition, self-similarities, and other techniques to compress arbitrary arrays of data, implemented as lookup tables. We showed the effectiveness of our method by implementing several nonlinear functions, constant coefficient multipliers, and Kolmogorov-Arnold networks at 12-bit resolution on FPGAs. In terms of throughput per LUT hardware cost, CompressedLUT was on average 3.30, 10.55, and 1.87 times better than conventional lookup table implementations for those three applications, respectively.}

\section*{Acknowledgments}
This material is based upon work supported in part by Cisco Systems, Inc. under grant number 00105407, and by the National Science Foundation under grant number PFI-TT 2016390.

\bibliographystyle{IEEEtran}
\bibliography{bibliography}

\end{document}